\documentclass[aps,preprintnumbers, preprint,superscriptaddress,nofootinbib]{revtex4}
\usepackage{amsmath}
\usepackage{graphicx}
\usepackage{amssymb}
\usepackage{color}
\usepackage{soul}

\newcommand{\beq}{\begin{equation}}
\newcommand{\eeq}{\end{equation}}
\newcommand{\bea}{\begin{eqnarray}}
\newcommand{\eea}{\end{eqnarray}}
\newcommand{\met}{\not{\!\!{ E}}_{T}}

\begin{document}

\title{Mono Vector-Quark Production at the LHC}

\author{Haiying Cai}
\email{hycai@pku.edu.cn}
\affiliation{Department of Physics, Peking University, Beijing 100871, China}

\begin{abstract}
Vector-like quark  is a common feature in many new physics models. We study the
vector-like quark production as an $s$-channel resonance at the Large Hadron Collider
for the vector-like quarks mainly mixing with the top- and bottom-quark in the Standard Model.
We emphasize  that the leptonic angular distribution can be used to discriminate various vector-like quark models.
\end{abstract}

\maketitle

\section{Introduction}

Vector-like quark  (vector-quark) exists in many models of new physics (NP) beyond the Standard Model (SM),
e.g. extra dimensional models, Little Higgs models and dynamical models, etc.
The vector-quark denoted by $\mathcal{Q}$ could mix with the SM quarks through Yukawa interaction.
In particular, the mixing of vector-quark  with the third generation quarks in the SM is quite common in the Little Higgs models~\cite{LH} and composite Higgs  Boson  models~\cite{CH}.
Several possibilities for their electroweak quantum numbers are listed in Ref.~\cite{delAguila} (see Table~\ref{quantum}). The phenomenology of the vector-quark production in hadron collisions has been studied either in the pair production \cite{PVQ} or in the  association production with a SM quark \cite{Barcelo,Atre}.
In this work we consider the mono vector-quark production as  an $s$-channel resonance
via  strong magnetic  $q$-$g$-$\mathcal{Q}$ couplings at the Large Hadron Collider (LHC).  Depending on whether the vector-quark would mix with SM chiral quarks, we  further focus on the decay channel that  vector-quark  decays into a top-quark plus a jet, i.e. $\mathcal{Q} \to  t g$  and  the electroweak decay channel $\mathcal{Q} \to  b W^+$.  The top-quark polarization can be measured through the  charged lepton angular distribution from top-quark decay inside the top-quark's rest frame and the polarization is sensitive to the chirality of the $g$-$t$-$\mathcal{Q}$ coupling. The $SU(2) \times U(1)_Y$ charged and neutral currents will be modified according to the quantum numbers of vector-quarks after adding Yukawa mixings, therefore the branch ratios for vector-quark decay to electroweak gauge bosons vary in  several vector-quark models.  Discovering the vector-quark signal and further measuring the top-quark polarization would help us to distinguish several new physics models.

The $q$-$g$-$\mathcal{Q}$ coupling is forbidden by the Ward Identity at the tree level. The interaction could be induced by the new physics (NP) at high energy scale ($\Lambda$). Rather than working in an ultra-violet completion theory, we adapt the approach of effective field theory (EFT) in this study. The vector-quark might carry different quantum numbers under the electroweak symmetry of the SM ($SU(2)_L \times U(1)_Y$).  It could be a weak isospin singlet, doublet or triplet. Below we list all the possible gauge invariant dimension-6 effective operators which describe the  strong  magnetic $g$-$q$-$\mathcal{Q}$ interactions for singlets, doublets and triplets respectively:
 \bea
  \mathcal{L}_{gq\mathcal{U}_{1},\mathcal{D}_1}&=& \frac{\kappa_{u, R}}{\Lambda^2}~\overline{q}_L \sigma^{\mu\nu} \lambda^A \mathcal{U}_{1 R} \tilde{\phi} G_{\mu\nu}^A
  + \frac{\kappa_{d, R}}{\Lambda^2} ~\overline{q}_L \sigma^{\mu \nu} \lambda^A \mathcal{D}_{1 R} \phi G_{\mu\nu}^A   \label{singlet} \\
  \mathcal{L}_{gq\mathcal{D}_2~~}&=& \frac{\kappa_{u, L}}{\Lambda^2} ~ \overline{\mathcal{D}}_{2L} \sigma^{\mu\nu} \lambda^A u_R \tilde{\phi} G_{\mu\nu}^A
 +\frac{\kappa_{d, L}}{\Lambda^2}~ \overline{\mathcal{D}}_{2L} \sigma^{\mu\nu} \lambda^A d_R \phi G_{\mu\nu}^A  \\
  \mathcal{L}_{g q\mathcal{D}_{X,Y}}&=& \frac{\kappa_{X, L}}{\Lambda^2} ~ \overline{\mathcal{D}}_{XL} \sigma^{\mu\nu} \lambda^A u_R \phi G_{\mu\nu}^A
 +\frac{\kappa_{Y, L}}{\Lambda^2} ~ \overline{\mathcal{D}}_{YL} \sigma^{\mu\nu} \lambda^A d_R \tilde{\phi} G_{\mu\nu}^A  \\
  \mathcal{L}_{g q \mathcal{T}_{X,Y}}&=& \frac{\kappa_{X, R}}{\Lambda^2}~ \overline{\mathcal{T}}_{XR}
 \sigma^{\mu\nu} \lambda^A \left(\phi \tau^I q_L\right) G_{\mu\nu}^A + \frac{\kappa_{Y, R}}{\Lambda^2}~ \overline{\mathcal{T}}_{YR} \sigma^{\mu\nu} \lambda^A ( \tilde{\phi} \tau^I q_L) G_{\mu\nu}^A,  \label{triplet}
\eea
where $\phi$ is the SM Higgs boson doublet, $G_{\mu\nu}^A$ is the field tensor of gluon and  $\Lambda$ is the cut off scale of our effective theory. Note that $\tau^I$ denotes the weak isospin in the basis of $(+,0,-)$. Table \ref{quantum} shows the detailed gauge quantum numbers for all the vector-quarks scenarios.
\begin{table}
\caption{Electroweak quantum numbers for vector-quark multiplets, which could mix with the SM quarks through the Yukawa interaction.\label{quantum}}
\begin{tabular}{c|ccccccc}
\hline\hline
$\mathcal{Q}^{(m)}$ & $\mathcal{U}_1$ & $\mathcal{D}_1$ &
$\mathcal{D}_2$ & $\mathcal{D}_X$ & $\mathcal{D}_Y$ & $\mathcal{T}_X$ &
$\mathcal{T}_Y$ \tabularnewline
\hline
& $U$ & $D$ &
$\left(\begin{array}{c}U\\D\end{array}\right)$ &
$\left(\begin{array}{c}X\\U\end{array}\right)$ &
$\left(\begin{array}{c}D\\Y\end{array}\right)$ &
$\left(\begin{array}{c}X\\U\\D\end{array}\right)$  &
$\left(\begin{array}{c}U\\D\\Y\end{array}\right)$
\tabularnewline \hline\hline
$SU(2)_L$ & 0 & 0 & 2 & 2 & 2 & 3 & 3
\tabularnewline \hline
$Y$ & 2/3 & -1/3 & 1/6 & 7/6 & -5/6 &2/3 & -1/3
\tabularnewline
\hline\hline
\end{tabular}
\end{table}

The paper is organized as follows. In the section~II we consider the
simplest scenario where the  vector-quark interacts  with the light
quark only through dimension-$6$ operators in the sense that
it  won't  mix with the Standard Model chiral quarks.The top-quark
is predominately left-handed polarized when it originates from a
gauge singlet or triplet vector-quark. On the other hand, it is
mainly right-handed polarized when it comes from a gauge doublet
vector-quark decay. Since top-quark is the only ''bare'' quark in
the SM, we consider the decay mode of $\mathcal{Q} \to t g$ to
utilize the top-quark polarization to distinguish the gauge
singlet/triplet and gauge doublet models. In the section~III  we
add the Yukawa interaction to couple the  vector-quark to the SM
chiral quark  such that fermions with the same quantum number can
mix. Electroweak precision measurements,  i.e. the $T$ parameter and
$Z \to b \bar b $,  are adopted to constrain the parameter space.
After that we proceed to analyze the channel of $ p p \to
\mathcal{Q}  \to b l^+  \nu$. We derive the leptonic angular
distribution with respect to the  gluon moving direction in the
center of mass  frame of initial partons,   which  can  be used to
discriminate the singlet/triplet  model from doublet model  under
certain assumptions.  We are going to show that the mixing pattern
for each vector quark scenario determines  their discovery
potentials in the LHC.

\section{Excited Quarks and Collider simulation}

\begin{figure}[]
\begin{center}
\includegraphics[scale=0.65]{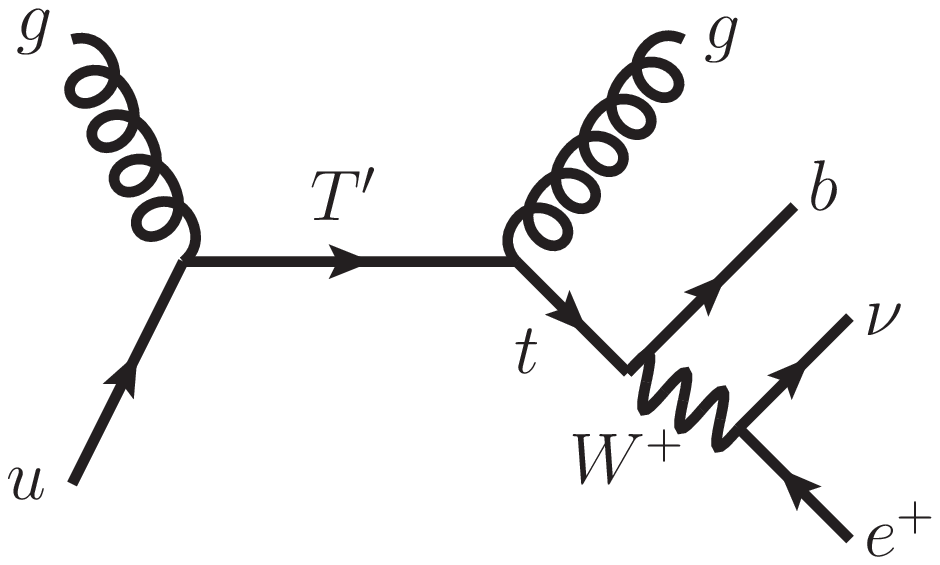}
\end{center}
\caption{Feynman diagram of the process of $u g \to \mathcal{Q}  \to  g t (\to b e^+ \nu_e)$.  \label{ugtg}}
\end{figure}

We are interested in the following process that the  mono produced  vector-quark  decays into a top-quark plus a jet ( see Fig.~\ref{ugtg} for  Feynman diagram ):
\beq
 u g \to \mathcal{Q}  \to t g, ~~ t \to b \ell^+ \nu,
\eeq
with $\mathcal{Q}$ carrying an electromagnetic charge $2/3$. In this section we assume that $\mathcal{Q}$ does not mix with the chiral top quark in the SM  such that they can only decay through  dimension-$6$  operators.  For clarity we only include the positively charged state in the following discussion.
Furthermore the leptonic decay mode of the top-quark is considered in the analysis because the top-quark spin is maximally correlated with the charged lepton $\ell^+$. The drawback is that neutrino is invisible. It escapes the detection and yields a signature of large missing energy ($\met$) at the collider. One has to determine the neutrino momentum to fully reconstruct the top-quark kinematics, which is the key for the top-quark polarization measurement.

After the spontaneous breaking of  electroweak symmetry, the effective operators described by Eq.~[\ref{singlet}- \ref{triplet}] induce an effective chromo-magnetic-dipole coupling of $g$-$u$-$\mathcal{Q}$ as follows:

\bea
L &\supset& \frac{{v}}{{{\sqrt{2}\Lambda ^2}}}  \bar q{\sigma ^{\mu \nu }}
\left( {{f_L}  {P_L} + {f_R}  {P_R}} \right) \mathcal{Q} {G_{\mu \nu }}(k) + h.c.
\eea
where $v=246~{\rm GeV}$ denotes the vacuum expectation value  for the Higgs field and
$P_L$, $P_R$ are the chiral projectors.  When we set the couplings $f_L$ and $ f_R $ to be of order one,  the cut off scale need to satisfy the condition of $\Lambda \gtrsim E $  so that the effective theory is sensible.  The effective operator will become important as the energy $E$ approaches the cut off scale, and particles with mass of  $\Lambda$ energy scale are likely to be produced, but  those particles should be irrelevant to the mono vector-quark production.  We further assume that the heavy quark $\mathcal{Q}$ interacts with all three up-type quarks in the SM ($u$, $c$, $t$) universally.  As we can see from its original dimension-$6$ form, for each vector-quark scenario the interaction should be either left handed or right handed depending on their representations  in the  $SU(2) $ gauge group, i.e.  $f_L $ and $f_R$ can not be present simultaneously.  This property makes the dimension-$6$ operators distinct from the dimension-$5$  operators written down in the excited quark model~\cite{excited}.  Similar FCNC  anomalous magnetic dimension-$6$ operators with the heavy quark $\mathcal{Q}$ replaced by the top quark are adopted to study the single top production~\cite{Tait, Hosch, Aguilar},  whose coupling bounds  are recently explored in the ATLAS experiments~\cite{Aad}.
\begin{figure}[tb]
\begin{center}
 \begin{minipage}{300 pt}
 \includegraphics[width =0.78 \hsize]{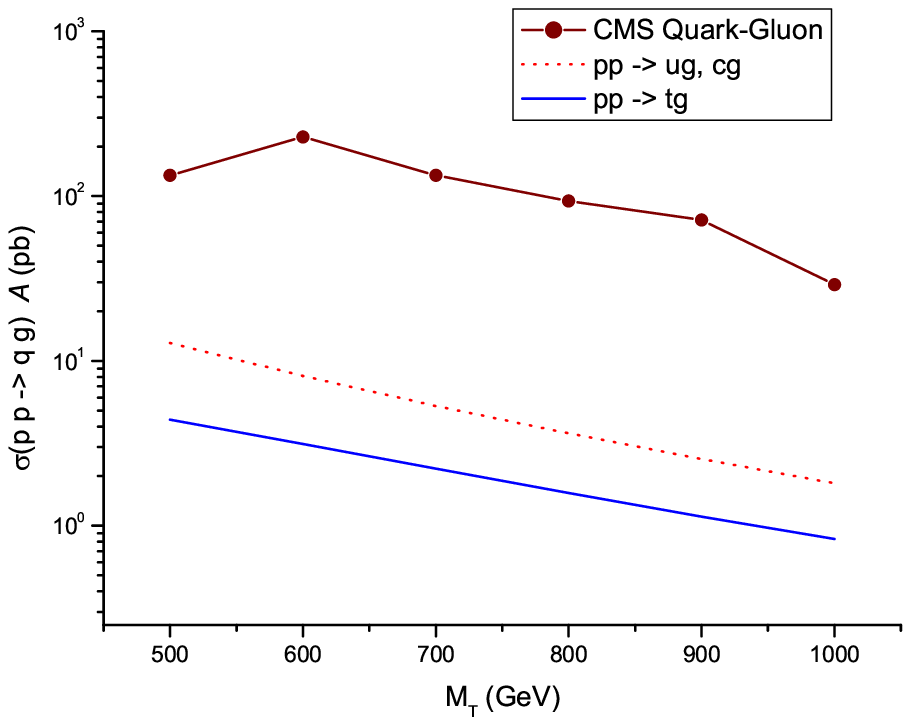}  
 \end{minipage}
  \begin{minipage}{300 pt}
 \includegraphics[width =0.78 \hsize]{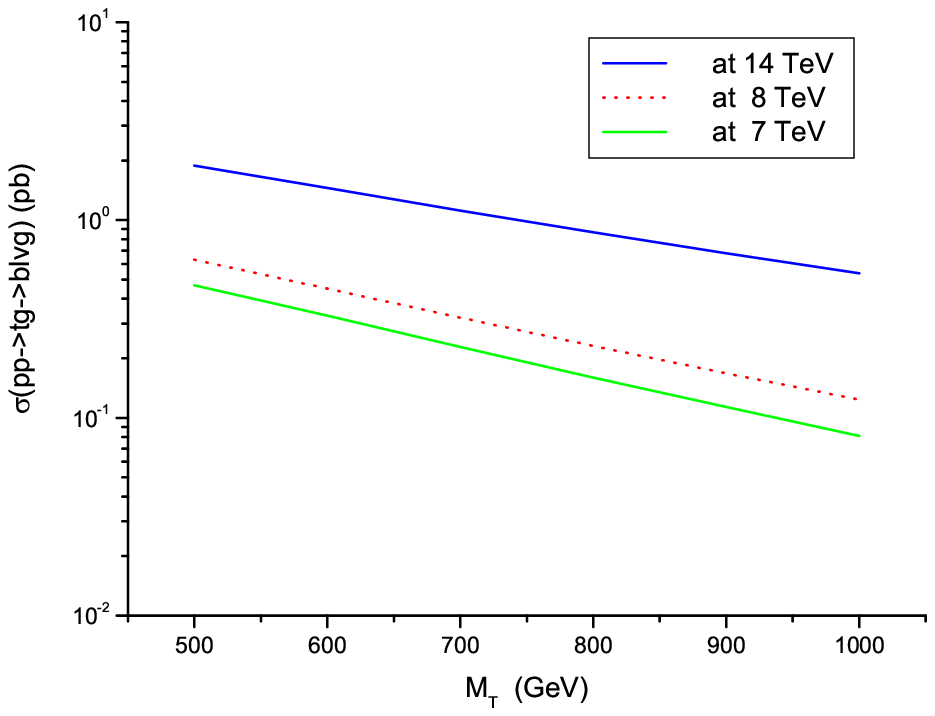}
 \end{minipage}
\end{center}
\caption{The upper panel plot is the inclusive cross section (in the unit of picobarn)  times  CMS acceptance  $A $  for the process  $ p p \to  T \to q g$ at  a 7~TeV LHC.  The red dotted line is for $u g $ and $c g$  final states via one excited $T$-quark,  the blue solid line is for $t g $ final state via one excited $T$-quark  and the dark red line is the CMS upper limit for quark-gluon final states.  The couplings are chosen as $f_L= 1.0$ and $f_R=0$ with the cut-off scale $\Lambda$  fixed to be 1.2  ~TeV.  The lower panel plot is the inclusive cross sections for the process $ p p \to T \to q g $ at 7  TeV,  8  TeV and 14 TeV LHC in the green solid line, red dotted line and blue solid  line respectively.  The couplings are chosen as $f_L= 0.5$ and $f_R=0$   with the same cut-off scale $\Lambda =  1.2$  ~TeV.
\label{xsec}}
\end{figure}

Since the excited heavy quark  $\mathcal{Q}$ has an electric charge $ 2/3 $, we denote it as  one  heavy  vector like $T$-quark in the following.  The single $T$-quark  production cross section at the LHC  is :
\beq
\sigma(qg \to T) =\frac{1}{S} \int dx_1 dx_2 \left\{ f_{q/P}(x_1) f_{g/P}(x_2) + (x_1 \leftrightarrow x_2)\right\} \delta(\hat{s} - m_T^2)  \frac{\pi\hat{s}v^2}{6\Lambda^4} (f_L^2 + f_R^2),
\eeq
where $f_{i/P}$ is the so-called parton distribution function (PDF) describing the possibility of finding a parton $i$ inside a proton with a momentum fraction $x_i$.  In the simplified model, the decay width is :
\bea
\Gamma (T \rightarrow q g ) = \frac{v^2 (f_L^2  + f_R^2  ) \left( m_T^2-m_q^2\right)^3}{6 \pi  \Lambda ^4 m_T^3},
\eea
which clearly displays that the branching ratio of $T\to t g$
increases with $m_T$  while the branch ratios of $ T \to u g, c g
$ decrease with $m_T $, and they all approach to $1/3$ in the
large $m_T$  limit. Since our effective operators  contribute to
the quark-gluon final states in the hadron collider, the upper
limit  of  the dijet cross section  puts some constraints on the
couplings $f_L $ and $f_R$ \cite{Han, cms}. The cross section
$\sigma$  times the CMS acceptance factor $A$ for  $p p \to  T \to
ug , cg $ at  a  7 TeV LHC  is compared with the corresponding CMS
upper limit for the quark-gluon final state  as shown in Fig.
\ref{xsec}.  As we can see that when we fix the cut-off  $\Lambda$ to  be $1.2$
TeV and require that $f_L <1.0, {\rm or}~   f_R <1.0 $,  the contribution from 
the effective operators is far below the CMS dijet constraint. 
The inclusive cross sections for the process of
$q g \to T \to b \ell^+ \nu g$ with $\ell^+ = e^+, \mu^+$ when the
center of mass energy  is 7  TeV, 8 TeV and  14  TeV are also
plotted in Fig.~\ref{xsec}, where  the couplings is chosen to be
$f_L=0.5$, $f_R=0$  (or $f_L=0$, $f_R=0.5$) and the  cut-off
$\Lambda$ is fixed to be 1.2~TeV.

The top-quark polarization can be best measured in its leptonic decay mode $t \to b \ell^+ \nu$.  The collider signature of interest to us is one charged lepton, two jets plus large missing energy, where the large missing energy originates from the invisible neutrino and one of the two jets is the gluon in association with top-quark production and the other one is  the b-quark from top-quark decay. The signature suffers from a few SM backgrounds as follows:
\begin{itemize}
\item Single-$t$ production:  single top-quark can be produced via the electroweak interaction in the Standard Model. It proceeds through the $s$-channel decay of a virtual $W$ ($q\bar{q}^\prime \to W^* \to t\bar{b}$), the $t$-channel exchange of a virtual $W$ boson ($b q \to t q^\prime$ , $b\bar{q}^\prime \to t \bar{q}$), and the associated production of a top-quark with two jets  $( q q^\prime , g q \to t j j ) $.  Note that $t j j $  includes the process of $g b \to t W^-$ and with $W^-$ subsequently decay into two jets. The single top plus one jet  process is the intrinsic background as it  yields exactly the same signature as the signal event. On the the hand, the  top quark associated production with two jets  process is the non-intrinsic background, since this process can only mimic the signal in the case that one of the two jets  produced in association with the top-quark escapes the calorimeter detection.
\item  $W + $ two jets  production processes:  among them are the  $Wbj$ production and the $Wb\bar b $ production which have the same signature as the signal events;  the other one is  $Wjj$ production process but it requires one of the two jets to fake the $b$-jet.  As to be shown later, it still contributes as the largest background even after including the small faking efficiency. Additional small background is from $WZ $ production with $Z$ gauge bosons  decay into two jets.
\item $t\bar{t}$ production: it is a non-intrinsic background because in order to mimic the signal, either the two jets from the $\bar{t}$ hadronic decay or the charged lepton from the $\bar{t}$ leptonic decay need to get lost in the detector.
\end{itemize}
Both the signals and backgrounds are generated by MadGraph/MadEvent~\cite{MG}. The CTEQ6L1 parton distribution functions~\cite{cteq} are used in this study. When generating the $Wjj$ backgrounds, we impose soft cuts on the transverse momentum of the both jets, i.e. $p_T(j)>5~{\rm GeV}$, to avoid the collinear singularity. Note that such soft cuts do not affect our analysis as we impose much harder cuts on both jets in the following analysis.  The inclusive cross section for the signal event  i.e.  $pp \to T\to t g$ with the following decays  $t\to b \ell^+ \nu$ ($\ell^+ = e^+,~\mu^+$),   is shown in the second column in Table~\ref{tg_tab}. For illustration we choose six benchmark masses for the $T$ quark and choose $f_L=0.5$ and $f_R=0$. The cut off scale is set to be  $\Lambda=1.2~{\rm TeV}$ throughout this work. Other results of different couplings can be easily obtained by rescaling:
\bea
\sigma_s(f_L,~f_R) = \sigma_s(f_L=0.5,~f_R=0) ~4~  (f_L^2 + f_R^2).
\eea
\begin{table}
\begin{center}
\caption{ Cross section (in the unit of fb) of the signal events for various $m_T$ at the 14~TeV LHC (upper panel) and of the SM backgrounds (lower panel). The model parameters are chosen as
$ f_L =0.5 $ , $ f_R =0$ and  $\Lambda = 1.2 ~ \mbox{TeV}$. }
\vspace*{0.5cm}
\begin{tabular}{ccccccc} \hline
$ m_{T} ~(\mbox{GeV}) \quad $ & no cuts  \quad &  basic  cuts \quad  & $ P^g_T> 200  ~\mbox{GeV} $ & $ H_T > 400 ~ \mbox{GeV} $ & $  \begin{array}{c}\Delta M_W < 10~\mbox{GeV}  \\ \Delta M_t < 10~\mbox{GeV}  \end{array} $
\\ \hline
$ 500 ~\mbox{GeV} $ & 1893.51 & 512.192 & 256.287 & 240.759 & 233.659 \\  \hline
$ 600 ~\mbox{GeV} $ & 1453.86 & 417.329 & 322.756 & 304.365 & 295.569 \\ \hline
$ 700 ~\mbox{GeV} $ & 1120. & 335.439 & 292.542 & 277.59 & 268.687    \\ \hline
$ 800 ~\mbox{GeV} $ & 869.751 & 276.362 & 255.054 & 242.617 & 235.703 \\  \hline
$ 900 ~\mbox{GeV} $ & 680.2 & 223.344 & 211.95 & 200.387 & 194.299    \\ \hline
$ 1000 ~\mbox{GeV} $ & 540.21 & 181.403 & 174.866 & 166.736 & 161.658  \\ \hline
\end{tabular}

\vspace*{0.5cm}

\begin{tabular}{ccccccc} \hline
 Backgrounds  \quad  & no cuts  &  basic  cuts   & $ P^g_T > 200  ~\mbox{GeV} $ & $ H_T > 400 ~ \mbox{GeV} $ & $  \begin{array}{c}\Delta M_W < 10 ~\mbox{GeV}  \\ \Delta M_t < 10 ~\mbox{GeV}  \end{array} $
\\ \hline
$ Wjj    $  & $1.8532 * 10^7  $ & $13476.5  $ & $637.501$ & 633.794  & $163.082$ \\ \hline
$ Wb \bar b  $ & $ 48054.  $ & $1191.74  $ & $13.215$ & 12.014  & $3.604$ \\ \hline
$ Wb j     $ & $ 2774.2  $     & $36.758  $ & $0.277$ & 0.277  & $0.139$ \\ \hline
$ W Z      $ & $544.13   $       & $99.494  $ & $0.449 $ & 0.422 & $0.027$ \\ \hline
$ t  j        $ & $29619.  $     & $4214.78  $ & $ 311.0 $& 294.71  & $281.38$ \\ \hline
$ t  j  j     $ & $28234. $      & $ 5522.57   $ & $364.219 $ & 290.81 & $237.166$ \\ \hline
$ t \bar b    $ &  $1025.1  $  & $247.649  $ & $12.945$  &12.004 & $9.718$ \\  \hline
$ t  \bar b  j  $ & $ 27745  $   & $ 975.237 $ & $40.230 $ & 31.907 & $26.358$ \\ \hline
$ t \bar {t} (\bar b \ell \nu)   $  &  $25431.  $ & $118.254  $& 3.178 & $1.272 $ & $0 . 0 $ \\  \hline
total  &  $1. 86954*10^7$ & $25883 $ & $1383.01 $ & 1277.21 & $ 721.474 $ \\  \hline
\end{tabular}
\end{center}
\label{tg_tab}
\end{table}

To simulate a realistic detection, we impose the event-selection cuts as follows:
\bea
&& p_T^b > 20~{\rm GeV},  ~  \left| {{\eta _b}} \right| < 2.5,  \nonumber \\
&& p_T^\ell > 20~{\rm GeV}, ~ \left| {{\eta _l}} \right| < 2.5, \nonumber \\
&& \Delta {R_{jj}} > 0.4, ~ \Delta {R_{jl}} > 0.4~,
\eea
where $\Delta R$($\equiv \sqrt{(\Delta \eta)^2 + (\Delta \phi)^2}$) is the separation between any two observable final state particles (not including neutrinos) , and  $\Delta \phi$ and $\Delta \eta$ are the separations in azimuthal angle and rapidity respectively.  In order to let those non-intrinsic backgrounds to mimic the signals,  additional veto cuts are demanded for  jets or leptons which are not detected (either falling into a large rapidity region or carrying a too small transverse momentum to be detected), 
\beq
{\rm veto~cuts:~~} p_T(j,\ell^\pm) < 10~{\rm GeV}~{\rm or}~|\eta(j,\ell^\pm)|>3.5~.
\eeq
with $\eta$ denoting the rapidity of the jets or leptons in the final state. The vetoing cuts are imposed at the same time with the basic cuts when we are performing the events selection  and the corresponding cross sections after the event selections are  presented in  the third column of Table~\ref{tg_tab}.  We model the detector resolution effects by smearing the final state energy according to
\beq
\frac{\delta E}{E} = \frac{\mathcal{A}}{\sqrt{E/{\rm GeV}}}\oplus \mathcal{B},
\eeq
where we take $\mathcal{A}=10(50)\%$ and $\mathcal{B}=0.7(3)\%$ for leptons (jets).
In addition we require that one jet is tagged as $b$-jet with a tagging efficiency of $50\%$. We also apply a mistagging rate for charm-quark $\epsilon_{c\to b}=10\%$ for $p_T(c)>50~{\rm GeV}$. The mistagging rate for a light jet is $\epsilon_{u,d,s,g\to b}=0.67\%$ for $p_T(j)<100~{\rm GeV}$ and $2\%$ for $p_T(j)>250~{\rm GeV}$. For $100~{\rm GeV}<p_T(j)<250~{\rm GeV}$, we linearly interpolate the fake rate given above.

\begin{figure}[h]
\begin{center}
\begin{tabular}{cc}
\includegraphics[width=0.45 \hsize]{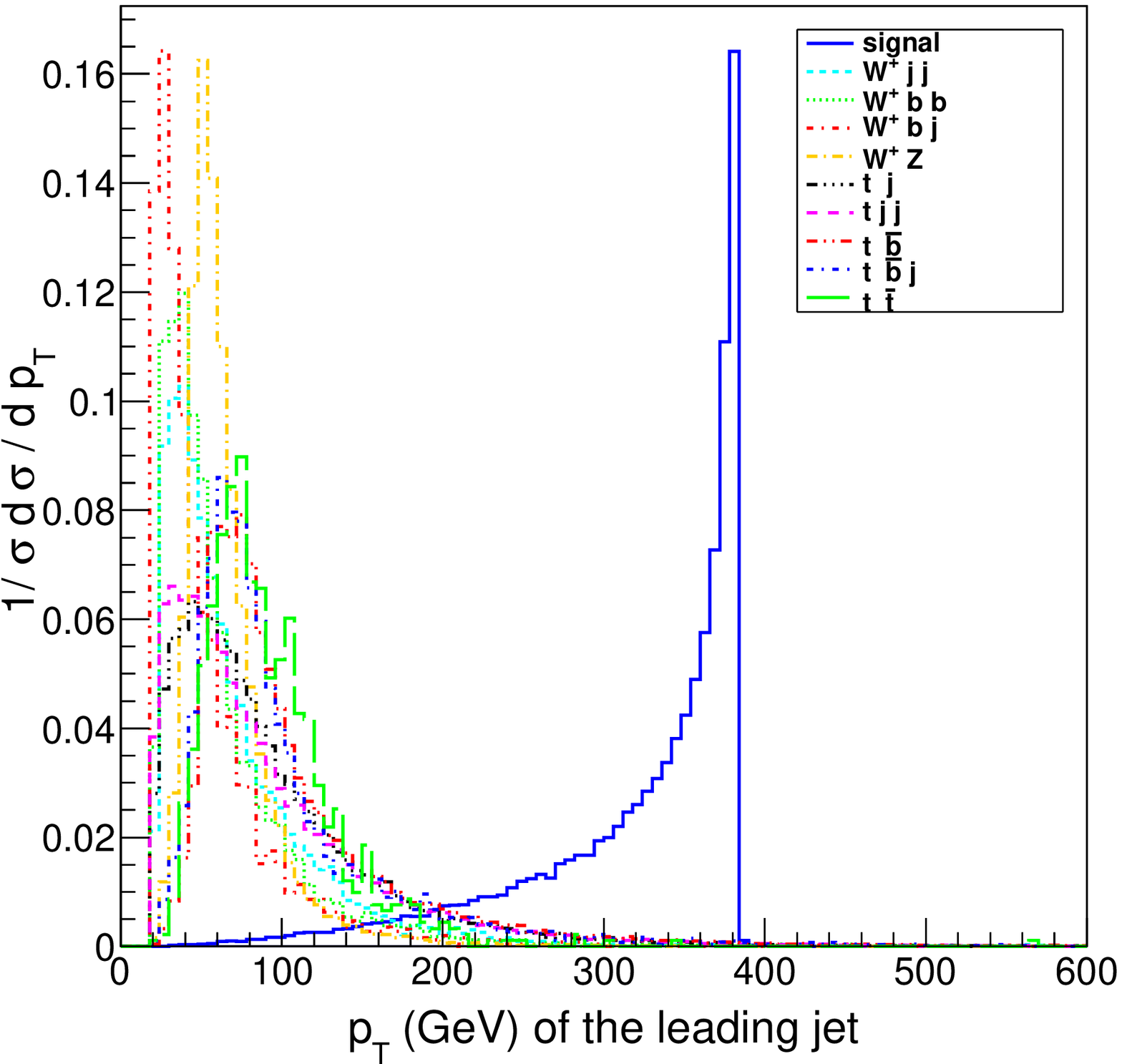} &
\includegraphics[width=0.45 \hsize]{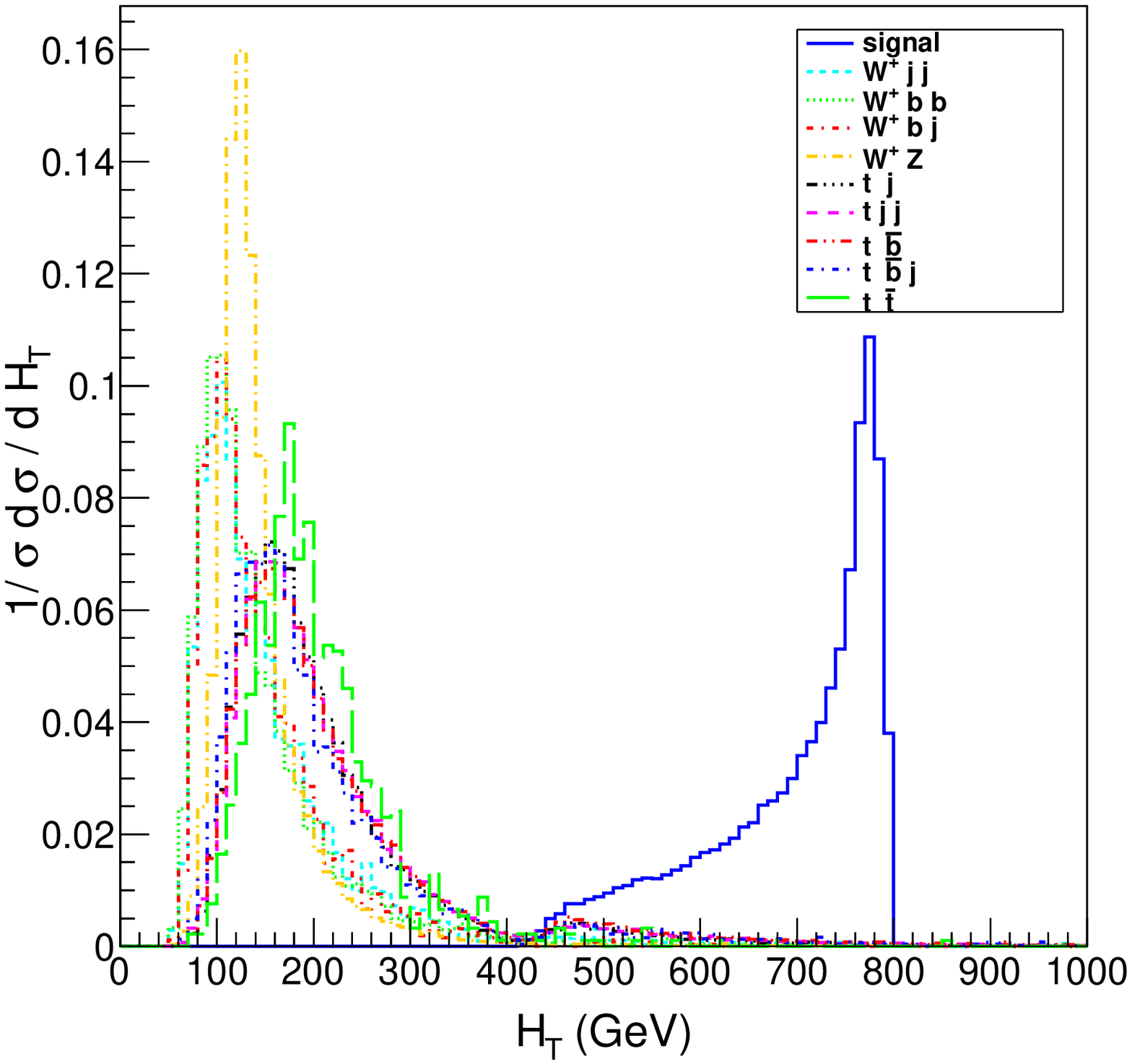}
\end{tabular}
\end{center}
\caption{The left panel is the $p_T$ distribution of the leading jet in the signal and background events  after the basic and veto selection cuts. The right panel is the $H_T$ distribution of all the final objects in signal and background events after the basic and veto selection cuts. In both plots we choose $m_T = 800 \mbox{GeV} $ and each distribution plot is normalized.  \label{ptht}}
\end{figure}
At this stage of the analysis, the background rate is one or two orders of magnitude larger than the signal rate. Moreover, the dominate background comes from the $Wjj$ process, followed by the single-top plus jets production and other small background processes. In order to study the efficient cuts that can significantly suppress the background rates while keeping most of the signal rates, we examine the $p_T$ distributions of the leading-$p_T$ jet as shown in Fig.~\ref{ptht}. The leading jet in the signal originates from the light jet produced in association with the top-quark. Owing to the large $m_T$, the leading jet peaks in the large $p_T$ region. On the contrary, the leading jets in the backgrounds, either from the QCD radiation or from the $W$-boson decay, tend to peak in the small $p_T$ region. Such a distinct difference enables us to impose a hard cut on the first leading jet,
\beq
p_T (j_{1st}) > 200~{\rm GeV},
\eeq
to suppress the huge backgrounds. In the fourth column of Table~\ref{tg_tab} we show the cross sections after the above cuts. This cut increases the signal-to-background ratio by a factor of  $17.2$  while keeping about  $92 \%$ of the signals for $m_T =800 \mbox{GeV}$. The biggest reduction in the background rate comes from the $Wjj$ production, but all the other backgrounds are reduced sizably as well.

One can further impose a hard cut on the $H_T$ variable, the scalar sum of the transverse momentum of all the measurable
objects in the final states i.e. $ H_T =  \sum p_T^i + \met $ with $i$ sum over the visible objects. The $H_T$ distribution of the signal events peaks in the large $H_T > 400 ~\mbox{GeV}$ region, while the one of the SM backgrounds is mostly located in the small $H_T < 400 ~ \mbox{GeV} $ region (see Fig \ref{ptht}).  But this cut does not influence both the signal and background too much after the hard $p_T (j_{1st}) > 200~{\rm GeV}$ cut .

\begin{figure}[tb]
\begin{center}
\includegraphics[width=0.4 \hsize]{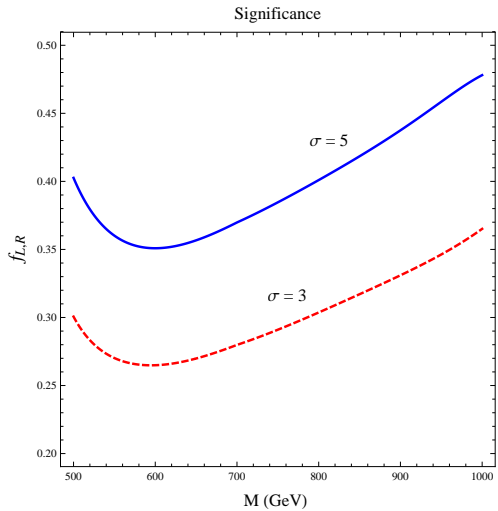}
\end{center}
\caption{Contour of the discovery potential of the signal at the 14~TeV LHC with an integrated luminosity of $1~{\rm fb}^{-1}$.  Both  the $3~\sigma $ exclusion limit $S / \sqrt {S+B} = 3$,  (red-dashed line)and the $5~\sigma$ discovery limit $S / \sqrt B = 5 $ (blue-solid line)  are shown in the figure.  \label{SGtg}}
\end{figure}

We are interested in measuring the top-quark polarization, which requires a full reconstruction of the top-quark kinematics. One confronts the invisible neutrino in the final state. Assuming the missing transverse momentum ($\met $)  comes  entirely from the neutrino, i.e.
\beq
 {p_{\nu}}(x) =  - \met(x), ~ {p_{\nu}}(y) =  - \met(y),
\eeq
the longitudinal momentum of the neutrino can be reconstructed by the on-shell condition of $W$-boson:
\bea
m_W^2 = (p_\ell+ p_\nu)^2.
\eea
The quadratic equation yields a two-fold solution. We only pick the real solutions and abandon the complex ones due to the $W$ gauge boson's width effect.   Since the $W$-boson comes from a top quark,  both solutions are  used to reconstruct the top quark mass,  so that we can  pick the one  which gives a mass closer to $173 ~ \mbox{GeV} $.   The neutrino momentum reconstruction is conducted after the large $p_T$ cut.
After neutrino reconstruction we are able to determine the kinematics of both the $W$-boson and top quark. Two mass-window cuts around $m_W$ and $m_t$ are imposed to optimize the signal events,
\beq
\Delta M_W = \left|m_{\ell \nu} - m_W\right| < 10~{\rm GeV}, \qquad \Delta M_t = \left|m_{\ell \nu b} - m_t\right|< 10~{\rm GeV}.
\eeq
Fig.~\ref{SGtg} plots discovery potential of the signal event in the plane of $m_T$ and $f_L$ ($f_R=0$) after the mass-window cuts.  We display both the $5 \sigma $ discovery limit and the $3 \sigma $ exclusion limit in the significance plot, which shows that it is promising to observe the single $T$-quark production at the LHC with an integrated luminosity of $1~{\rm fb}^{-1}$.

\begin{figure}[h]
\begin{center}
\includegraphics[width=0.45 \hsize]{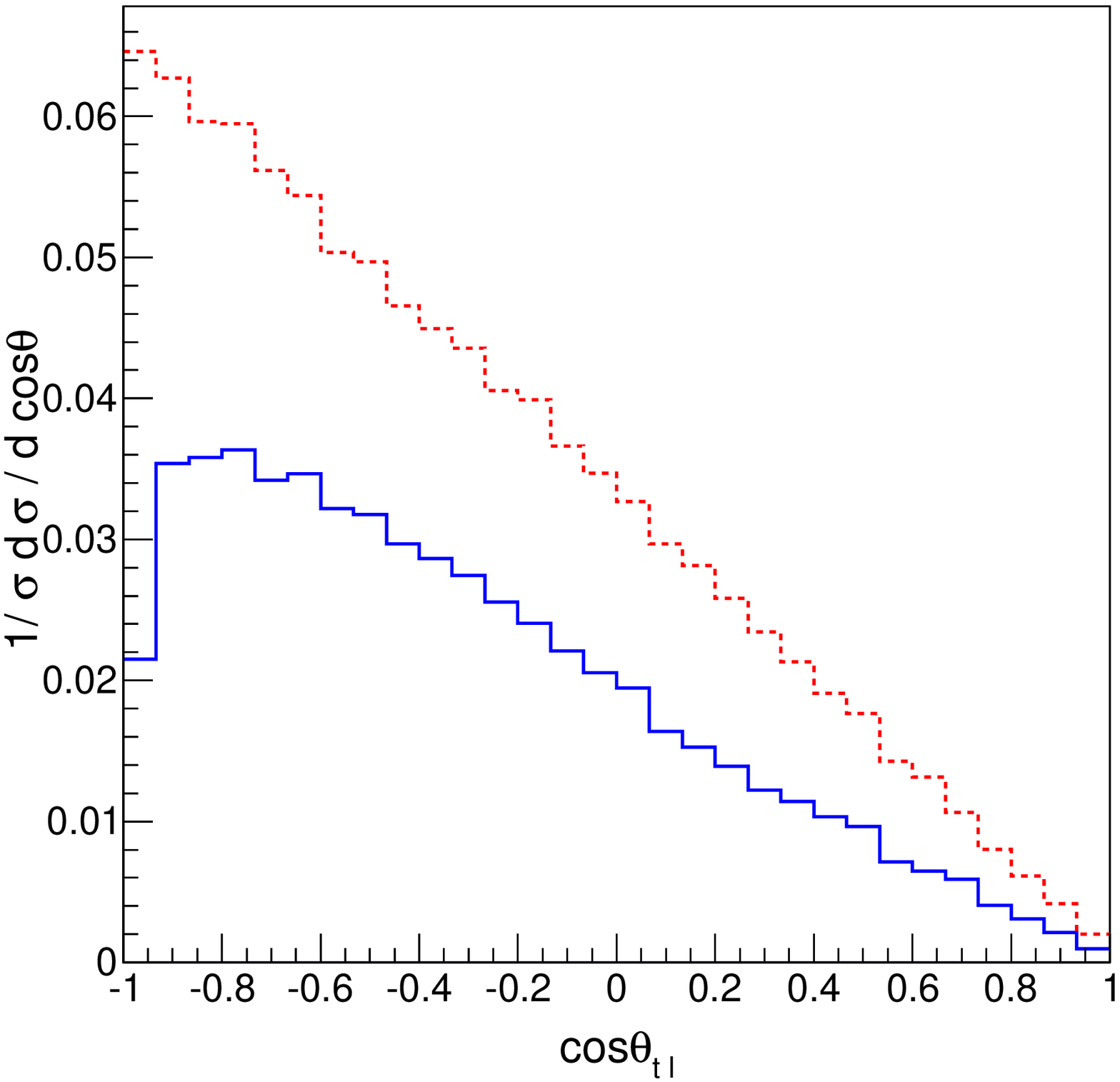}
\includegraphics[width=0.45 \hsize]{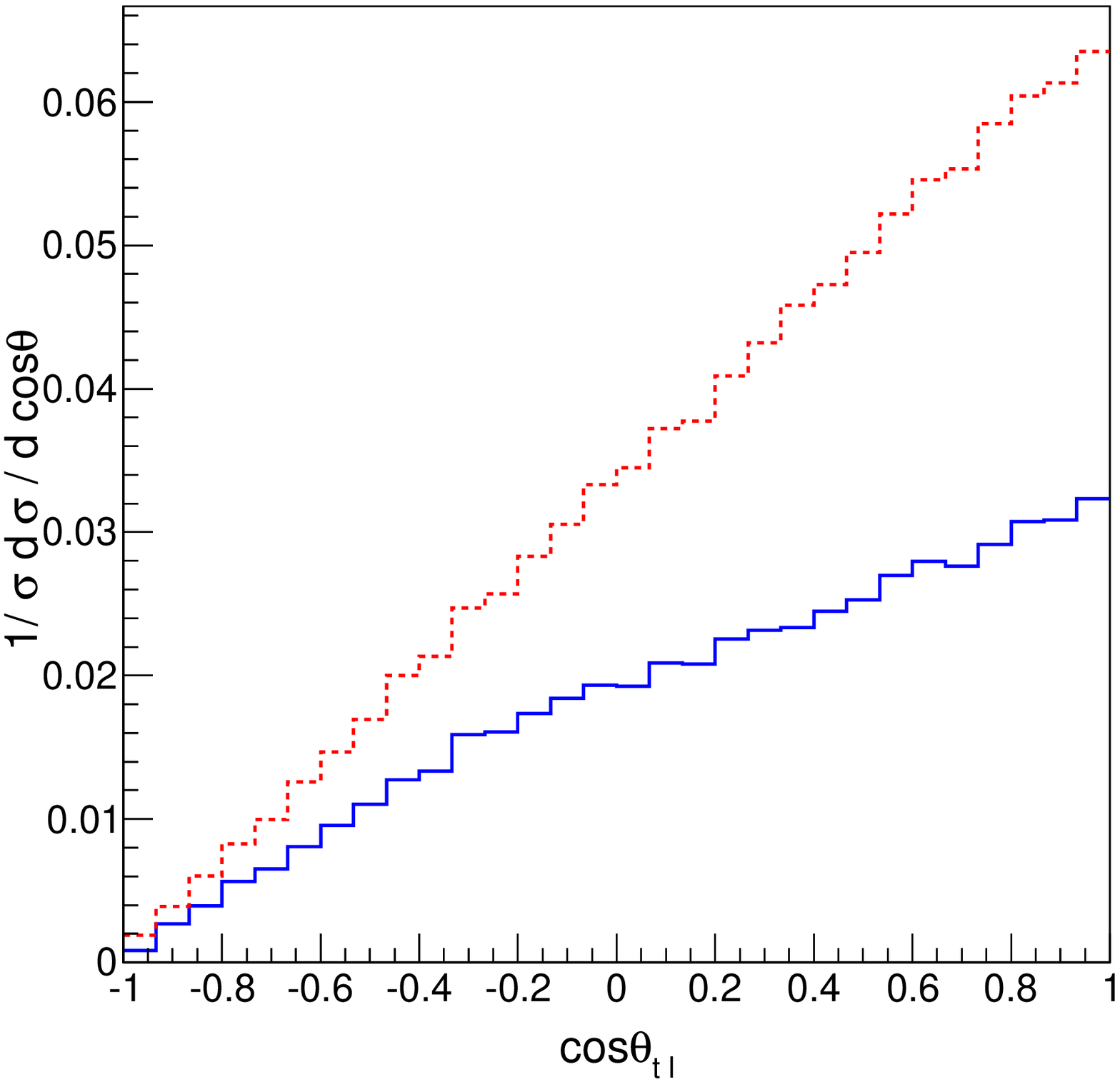}
\end{center}
\caption{The $\cos\theta$ distributions of the signal with no cut (dashed line) and after the mass-window cuts (solid line): (a) the singlet/triplet $T$-quark, (b) the doublet $T$-quark .
\label{costheta}}
\end{figure}

The behavior of the  angle between the lepton in the rest frame of the top quark and  the top quark moving
direction  in the center of mass frame can be factorized out  and  described by the following equation  \cite{kuhn,Parke}:
\bea
\frac{1}{\sigma}\frac{d \sigma } { d \cos \theta_{\ell} } = \frac{1}{2} (1 + a \cos \theta_{\ell} ),
\eea
with $a =1 $ for the pure left-handed heavy top quark and $a =-1 $ for the pure right-handed heavy top quark.
Figure~\ref{costheta} displays the $\cos\theta_\ell$ distributions with no cut and  after imposing the mass-window cut.

\section{Vector-quark and SM quark mixing}

In the previous discussion we assume the vector-quarks do not mix with the SM quark at the tree level. However one could write down gauge-invariant Yukawa interactions between the vector-quarks and the SM quarks as follows:
\bea
\mathcal{L}_{Y, SM} &=&  - {y_u}{{\bar q}_L}{H^c}{u_R} - {y_d}{{\bar q}_L}H{d_R}   \\
\mathcal{L}_{\mathcal{U}_1, \mathcal{D}_1}&=&  - {\lambda _u}{{\bar q}_L}{H^c}{ \mathcal{U}_{1 R}} - {\lambda _d}{{\bar q}_L}{H}{ \mathcal{D}_{1 R}}  - M{\bar{ \mathcal{U}}_L}{\mathcal{U}_R}  - M{\bar{ \mathcal{D}}_L}{\mathcal{D}_R}  \\
 \mathcal{L}_{ \mathcal{D}_2 ~~ }&=& - {\lambda _u}{\bar {\mathcal{D}}_{2L}}{H^c}{u_R} - {\lambda _d}{\bar {\mathcal{D}}_{2L}}H{d_R}  - M{\bar{ \mathcal{D}}_{2L}}{\mathcal{D}_{2R}}  \\
\mathcal{L}_{\mathcal{D}_{X,Y}} &=&  - {\lambda _u}{\bar {\mathcal{D}}_{XL}}{H}{u_R}- {\lambda _d}{\bar {\mathcal{D}}_{YL}}{H^c}{d_R}  - M{\bar{\mathcal{D}}_{XL}}{\mathcal{D}_{XR}}  - M{\bar{\mathcal{D}}_{YL}}{\mathcal{D}_{YR}} \\ 
\mathcal{L}_{\mathcal{T}_{X,Y}} &=&  - {\lambda _u}{{\bar q}_L}{\tau ^a}{H^c} \mathcal{T}_{XR}^a - {\lambda _d}{{\bar q}_L}{\tau ^a}H \mathcal{T}_{YR}^a- M{\bar {\mathcal{T}}_{XL}}{\mathcal{T}_{XR}} - M{\bar{ \mathcal{T}}_{YL}}{\mathcal{T}_{YR}}  ~ 
\eea
The Yukawa interactions generate  mixing between the SM quarks and the vector-quarks at the tree level after the spontaneous symmetry-breaking.  The singlet vector-quark and the triplet vector-quark exhibit a similar mixing pattern, while the doublet vector-quark has a different mixing pattern.
For simplicity we consider the scenario that the vector-quarks mix only with the third-generation quark in the SM .
Consider one pair of  vector-quark $T^\prime_{L,R} $ or one pair of vector-quark $ B^\prime_{L,R} $ which will mix with the chiral  top-quark $t^\prime_{L,R}$ or  the chiral bottom quark $b^\prime_{L,R} $ in the following way :
\bea
t^\prime_{L,R} &=& \cos \theta_{L,R}^u  t_{L,R} + \sin \theta_{L,R}^u T_{L,R},  \\
T^\prime_{L,R} &=& -\sin  \theta_{L,R}^u  t_{L,R} + \cos \theta_{L,R }^u T_{L,R},
\eea
\bea
b^\prime_{L,R} &=& \cos \theta_{L,R}^d  b_{L,R} + \sin \theta_{L,R}^d B_{L,R},  \\
B^\prime_{L,R} &=& -\sin  \theta_{L,R}^d  b_{L,R} + \cos \theta_{L,R}^d B_{L,R},
\eea
where $t_{L,R}$ and $T_{L,R}$ label the physical top-quark and  heavy vector-quark respectively.  Define two parameters $ x_t  = \lambda_u v/ \sqrt 2 $ and $ x_b = \lambda_d v/ \sqrt 2 $ and those mixing angles can  be calculated by diagonalizing the mass matrix. For the singlet/triplet vector-quark we obtain,
\bea
\sin {\theta_{u(d)}^L} &=& \frac{{M{x_{t(b)}}}}{{\sqrt {{{({M^2} - m_{t(b)}^2)}^2} + {M^2}x_{t(b)}^2} }}, \\
\sin {\theta _{u(d)}^R} &=& \frac{{{m_{t(b)}}{x_{t(b)}}}}{{\sqrt {{{({M^2} - m_{t(b)}^2)}^2} + {M^2}x_{t(b)}^2} }}.
\eea
while for the doublet vector-quark we have,
\bea
\sin {\theta _{u(d)}^L} &=& \frac{{{m_{t(b)}}{x_{t(b)}}}}{{\sqrt {{{({M^2} - m_{t(b)}^2)}^2} + {M^2}x_{t(b)}^2} }}, \\
\sin {\theta _{u(d)}^R} &=& \frac{{M{x_{t(b)}}}}{{\sqrt {{{({M^2} - m_{t(b)}^2)}^2} + {M^2}x_{t(b)}^2} }}.
\eea
Since the mixing would inevitably modify the $W$-$t$-$b$ and $Z$-$b$-$b$ couplings in the SM, the electroweak precision measurements at the low energy would severely constrain the mixing parameters~\cite{Cacciapaglia}.   Before proceeding with the further analysis, we are going to consider the low energy precision test in  several vector-quark models and put bounds on the parameter space using current experimental constraints. When we fix the parameter $x_b$, which describes the mixing of the bottom-quark and the vector-quark, the stringent constraint in the $(M, x_t)$ parameter space of the singlet and doublet vector-quark models is from the $T$-parameter, while it is from the $Z$-$b$-$b$ coupling for the triplet vector-quark model. The mixing of vector-quark with SM chiral quark modifies the charged current of $W^\mu_1$ gauge bosons and the neutral current of  $W^\mu_3$ gauge bosons simultaneously and the mass splitting among the fermions  will contribute  to  $T $ parameter through the vacuum polarization of  the gauge bosons \cite{peskin, Lavoura}. For the singlet vector-quark  $\mathcal{U}_1$, two types of doublet vector-quark $ \mathcal{D}_X$ and  $\mathcal{D}_2$ and the triplet vector-quark  $\mathcal{T}_X$,  their specific contributions to $T$ parameter  are  formulated in Eq.~[\ref{Tsg}-\ref{T3a}]:
\bea
\Delta T_{\mathcal{ U}_1}  &=&  \frac{3}{16\pi s^2 _W c^2 _W} \left [ \sin^2 \theta_u^{L  }{\theta _ + }({y_T},{y_b})
- \sin^2 \theta_u^{L  } {\theta _ + }({y_t},{y_b})  -   \cos^2 \theta_u^{L  } \sin^2 \theta_u^{L  } {\theta _ + }({y_T},{y_t})  \right]  \label{Tsg}
\\    \nonumber \\  
 \Delta T_{\mathcal{D}_X} &= & \frac{3}{16\pi s^2 _W c^2 _W}\left[ \sin^2  \theta_u^{L }{\theta _ + }({y_T},{y_b})  - \sin^2 \theta_u^{L  } {\theta _ + }({y_t},{y_b})
  \nonumber  \right. \\  &+ &  \left.   ( \sin^2  \theta_u^{L } + \sin^2  \theta_u^{R }  ) {\theta _ + }(y_t, y_X )
+  (\cos^2  \theta_u^{L } + \cos^2  \theta_u^{R }  ) {\theta _ + }(y_T ,  y_X )  \nonumber \right.  \\ &+ &
 \left.       2   \sin \theta_u^{L } \sin \theta_u^{R }  {\theta _ - }(y_t , y_X) +   2   \cos \theta_u^{L } \cos \theta_u^{R }   {\theta _ - }(y_T , y_X)   \nonumber \right. \\ &-&
\left.( 4 \cos^2 \theta_u^L \sin^2 \theta_u^L  + \cos^2 \theta_u^{R } \sin^2 \theta_u^{R } ) {\theta _ + }(y_t,y_T)  \nonumber \right. \\ &-&
\left. 4 \cos \theta_u^L \sin \theta_u^L   \cos \theta_u^{R } \sin \theta_u^{R }  {\theta _ - }(y_t,y_T) \right]    \label{Tna} 
\\ \nonumber  \\   
 \Delta T_{\mathcal{D}_2} &= & \frac{3}{16\pi s^2 _W c^2 _W}\left[(  \cos^2 (\theta_u^L - \theta_d^L)+ \sin^2 \theta_u^{R } \sin^2 \theta_d^{R } -1 )
 {\theta _ + }(y_t , y_b)   \nonumber  \right. \\  &+& \left. (  \cos^2 (\theta_u^L - \theta_d^L)+ \cos^2 \theta_u^{R } \cos^2 \theta_d^{R } ) {\theta _ + }(y_T , y_B)   \nonumber  \right. \\  &+&
\left.  (\sin^2 (\theta_u^L- \theta _d^L) +  \cos^2 \theta_d^{R } \sin^2 \theta_u^{R } ) {\theta _ + }(y_t,y_B)    \nonumber  \right. \\  &+&
\left.  (\sin^2 (\theta_u^L- \theta _d^L) +  \sin^2 \theta_d^{R } \cos^2 \theta_u^{R } )  {\theta _ + }(y_T,y_b)  \nonumber \right.  \\ &+ &
 \left.      2   \cos (\theta_u^L - \theta_d^L) (\sin \theta_u^{R } \sin \theta_d^{R }   {\theta _ - }(y_t , y_b) + \cos \theta_u^{R } \cos \theta_d^{R }   {\theta _ - }(y_T , y_B)  )   \nonumber \right.  \\ &+ &
 \left.      2  \sin (\theta_u^L- \theta _d^L)  ( \cos \theta_d^{R } \sin \theta_u^{R }   {\theta _ - }(y_t , y_B)  - \sin \theta_d^{R } \cos \theta_u^{R }  {\theta _ - } (y_T , y_b)  )   \nonumber \right.  \\ & - &
 \left.      \cos^2  \theta_u^{R } \sin^2  \theta_u^{R } {\theta _ + }(y_t,y_T)  -\cos^2  \theta_d^{R } \sin^2  \theta_d^{R } {\theta _ + }(y_b,y_B)     \right]    \label{Tdb} \eea
\bea
\Delta T_{\mathcal{ T}_X}  &=&  \frac{3}{16\pi s^2 _W c^2 _W} \left [( (\cos \theta_d^L  \cos \theta_u^L  + \sqrt 2 \sin \theta_d^L \sin \theta_u^L )^2 + 2 \sin^2 \theta_d^R  \sin^2 \theta_u^R -1 )  {\theta _ + }({y_t},{y_b})
\nonumber \right. \\ &+& \left.  2 \sqrt 2 (\cos \theta_d^L  \cos \theta_u^L  + \sqrt 2 \sin \theta_d^L \sin \theta_u^L) \sin \theta_d^R  \sin \theta_u^R  {\theta _ - }({y_t},{y_b})
\nonumber \right. \\ &+& \left. ( (\cos \theta_u^L \sin \theta_d^L  - \sqrt 2 \cos \theta_d^L \sin \theta_u^L  )^2 + 2  \cos^2 \theta_d^R  \sin^2 \theta_u^R   )  {\theta _ + }({y_t},{y_B})
\nonumber \right. \\ &-& \left.  2 \sqrt 2 ( \cos \theta_u^L \sin \theta_d^L  - \sqrt 2 \cos \theta_d^L \sin \theta_u^L) \cos \theta_d^R  \sin \theta_u^R {\theta _ - }({y_t},{y_B})
\nonumber \right. \\ &+& \left. ( ( - \sqrt 2  \cos \theta_u^L  \sin \theta_d^L + \cos \theta_d^L \sin \theta_u^L)^2 + 2  \cos^2 \theta_u^R   \sin^2 \theta_d^R   )  {\theta _ + }({y_T},{y_b})
\nonumber \right. \\ &-& \left.  2 \sqrt 2 (- \sqrt 2  \cos \theta_u^L  \sin \theta_d^L + \cos \theta_d^L \sin \theta_u^L )  \cos \theta_u^R   \sin \theta_d^R {\theta _ - }({y_T},{y_b})
\nonumber \right. \\ &+& \left. ( (\sin \theta_d^L  \sin \theta_u^  L  +  \sqrt 2 \cos \theta_d^L  \cos \theta_u^L )^2 + 2  \cos^2 \theta_d^R  \cos^2 \theta_u^R )  {\theta _ + }({y_T},{y_B})
\nonumber \right. \\ &+& \left.  2 \sqrt 2 (\sin \theta_d^L  \sin \theta_u^ L  +  \sqrt 2 \cos \theta_d^L  \cos \theta_u^L)  \cos \theta_d^R  \cos \theta_u^R   {\theta _ - }({y_T},{y_B})
\nonumber \right. \\ &+& \left. ( 2  \sin^2 \theta_u^L +  2 \sin^2 \theta_u^R )  {\theta _ + }({y_t},{y_X})  +  4  \sin \theta_u^L    \sin \theta_u^R   {\theta _ - }({y_t},{y_X})
\nonumber \right. \\ &+& \left. ( 2  \cos^2 \theta_u^L +  2 \cos^2 \theta_u^R )  {\theta _ + }({y_T},{y_X}) +   4  \cos \theta_u^L  \cos \theta_u^R   {\theta _ - }({y_T},{y_X})
\nonumber \right. \\ &-& \left.    (\cos^2 \theta_d^L \sin^2 \theta_d^L+ 4   \cos^2 \theta_d^R  \sin^2 \theta_d^R)  {\theta _ + }({y_b},{y_B})
\nonumber \right. \\ &-& \left.    4 \cos \theta_d^L \sin \theta_d^L   \cos \theta_d^R  \sin \theta_d^R  {\theta _ - }({y_b},{y_B}) -  \cos \theta_u^{L ~  2} \sin \theta_u^{L ~  2} {\theta _ + }({y_T},{y_t})   \right]  \label{T3a}
\eea
where $y_i =  m_i^2 / m_Z^2  $ is the dimensionless fermion mass squared rescaled by the inverse mass squared of the $Z$ gauge bosons. The SM constributions are substracted and I have  checked that all  the divergences are cancelled due to the  unitary  property of mixing matrix and the relation between those  mass eigenstates.  Since the $T$ parameter measures the effect  of  custodial symmetry breaking,  the functions $\theta_+ (y_1, y_2 )$ and $\theta_-(y_1, y_2)$  are zeroes when $y_1 = y_2 $ and they are  defined by the Ref.~\cite{Lavoura} :
\bea
{\theta _ + }({y_1},{y_2}) & =  &  {y_1} + {y_2} - \frac{{2{y_1}{y_2}}}{{{y_1} - {y_2}}}\log \frac{{{y_1}}}{{{y_2}}} \\
{\theta _ - }({y_1},{y_2})  &= & 2\sqrt {{y_1}{y_2}} \left( {\frac{{{y_1} + {y_2}}}{{{y_1} - {y_2}}}\log \frac{{{y_1}}}{{{y_2}}} - 2} \right)
\eea
\begin{figure}
\begin{center}
\includegraphics[width=0.35 \hsize]{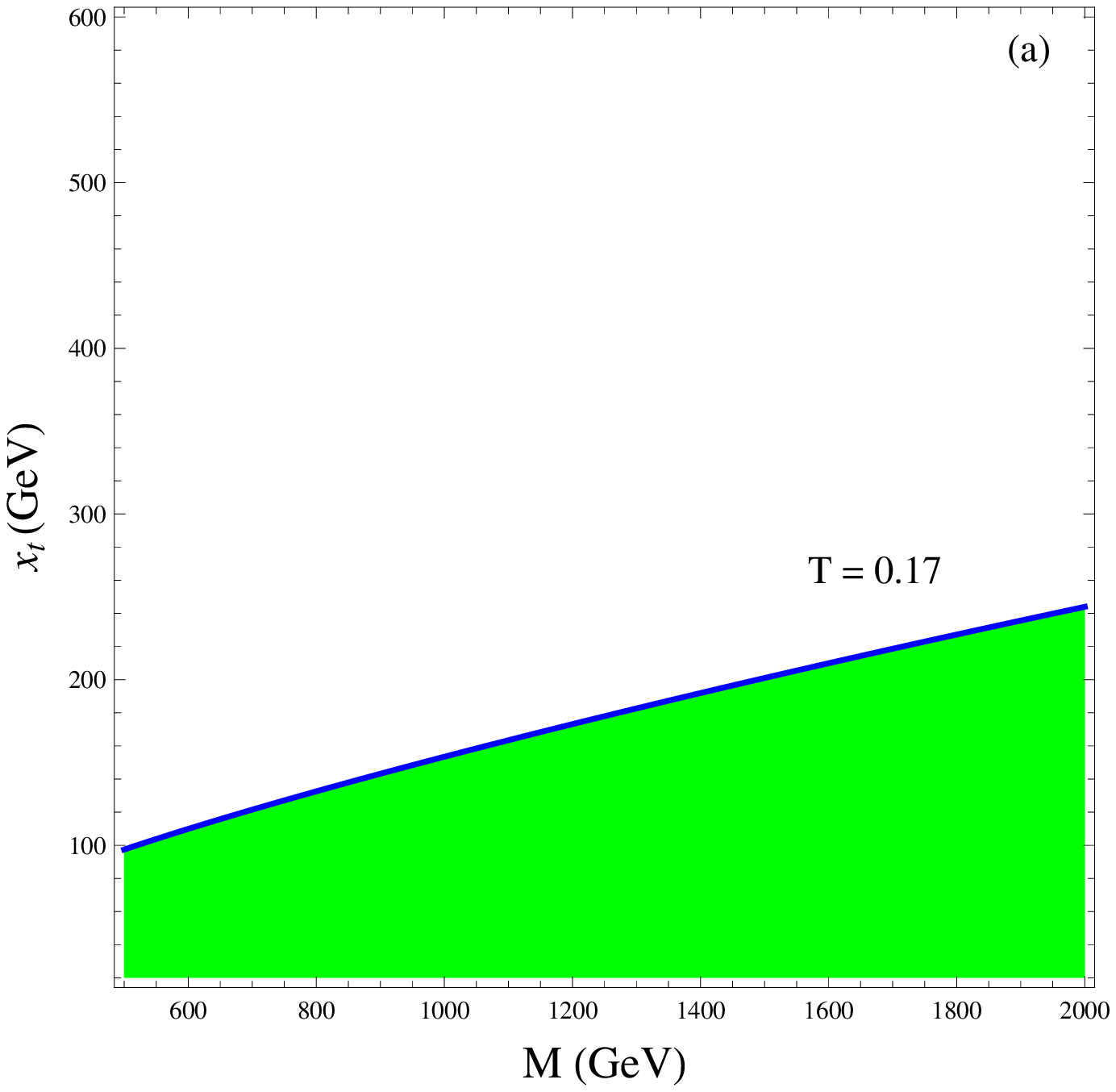}  \hspace{0.2 cm}
\includegraphics[width=0.35 \hsize]{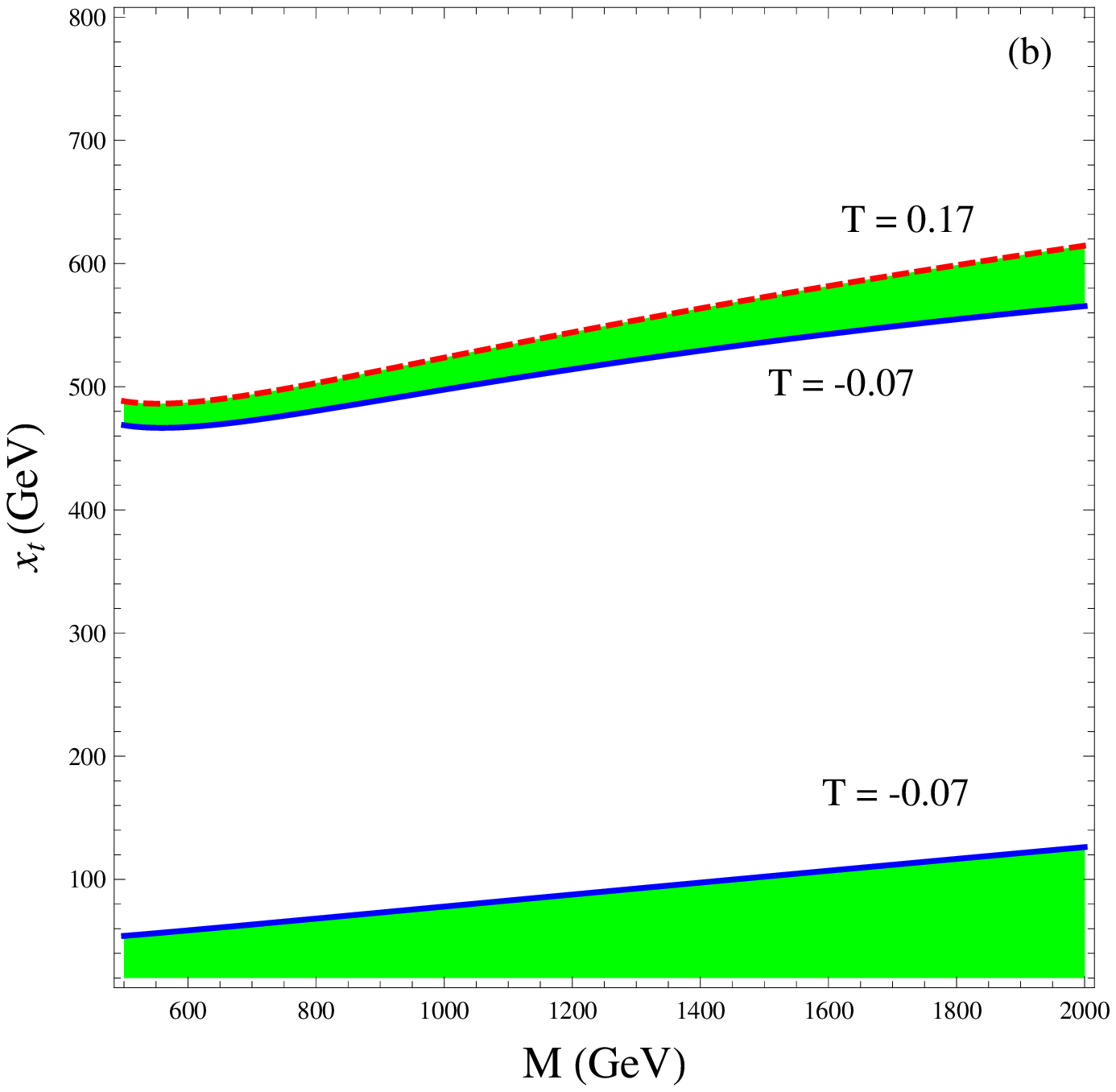}
\includegraphics[width=0.35 \hsize]{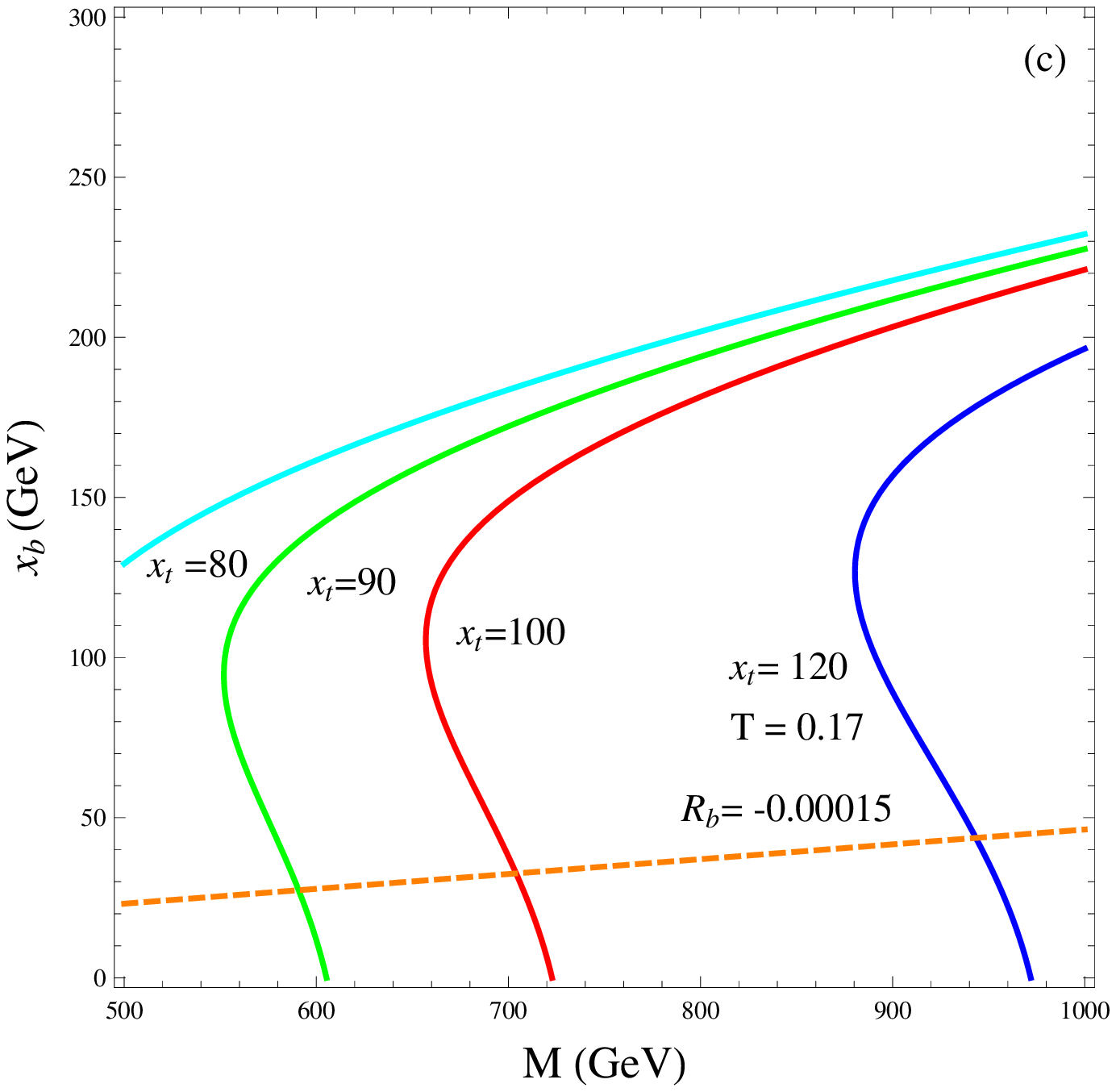}   \hspace{0.2 cm}
\includegraphics[width=0.35 \hsize]{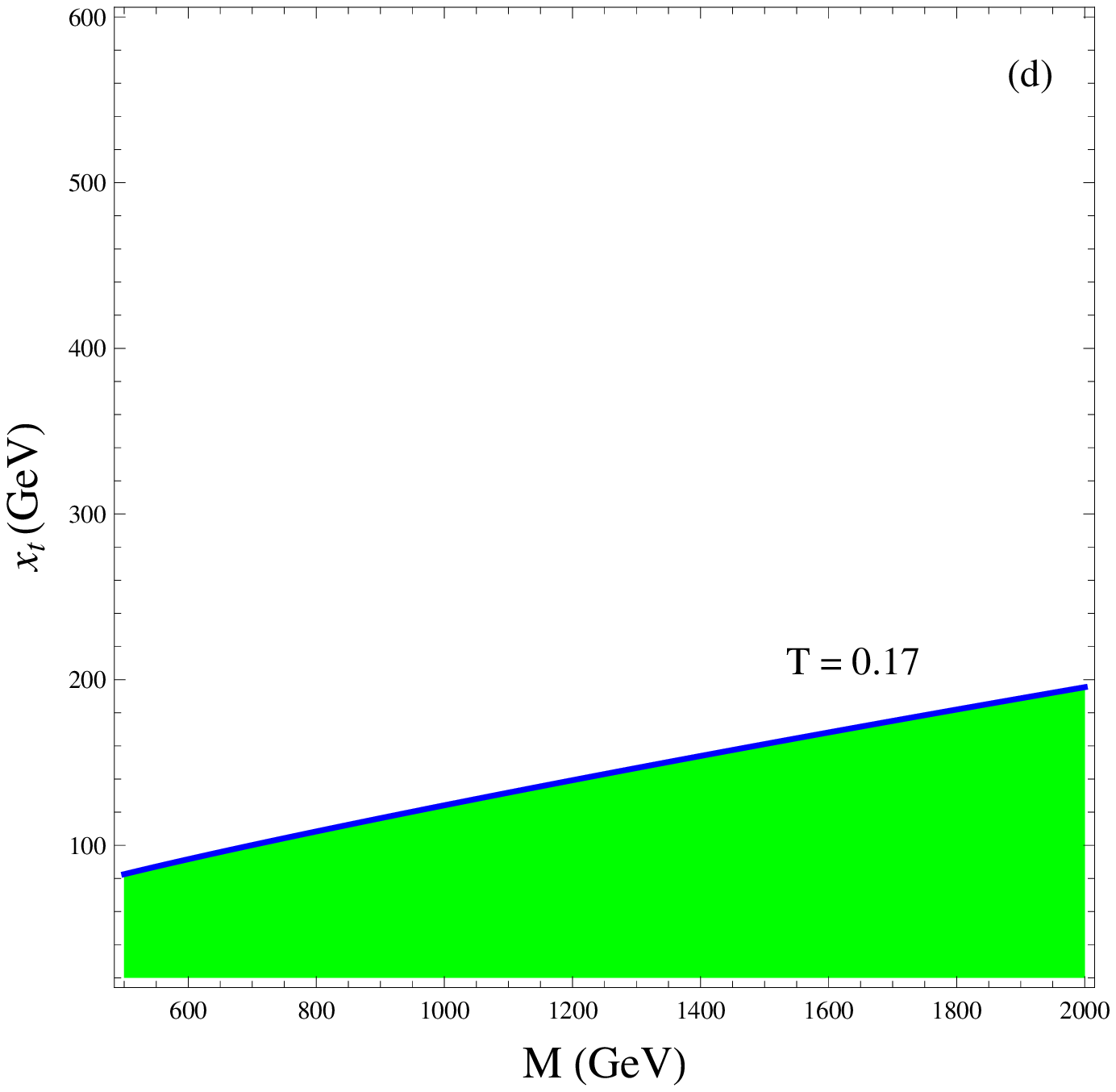}\end{center}
\caption{ (a) $T $ parameter  constraint  for $M$  and $x_t$ in the singlet $\mathcal{U}_1$ model ; (b) $T$ parameter constraint for $M$ and $x_t$  in  the doublet $\mathcal{D}_X$ model; (c) $T $ parameter  and $R_b $ constraints for  $M$ and $x_b$  in the doublet $\mathcal{D}_2$ model with $x_t$ fixed to be $80, 90, 100, 120 $ GeV ;  (d) T parameter constraint for $M$ and $x_t $ in the doublet $\mathcal{D}_2$ model with $x_b$ fixed to be $20 $ GeV.~\label{TPara}}
\end{figure}
\begin{figure}
\begin{center}
\includegraphics[width=0.4\hsize]{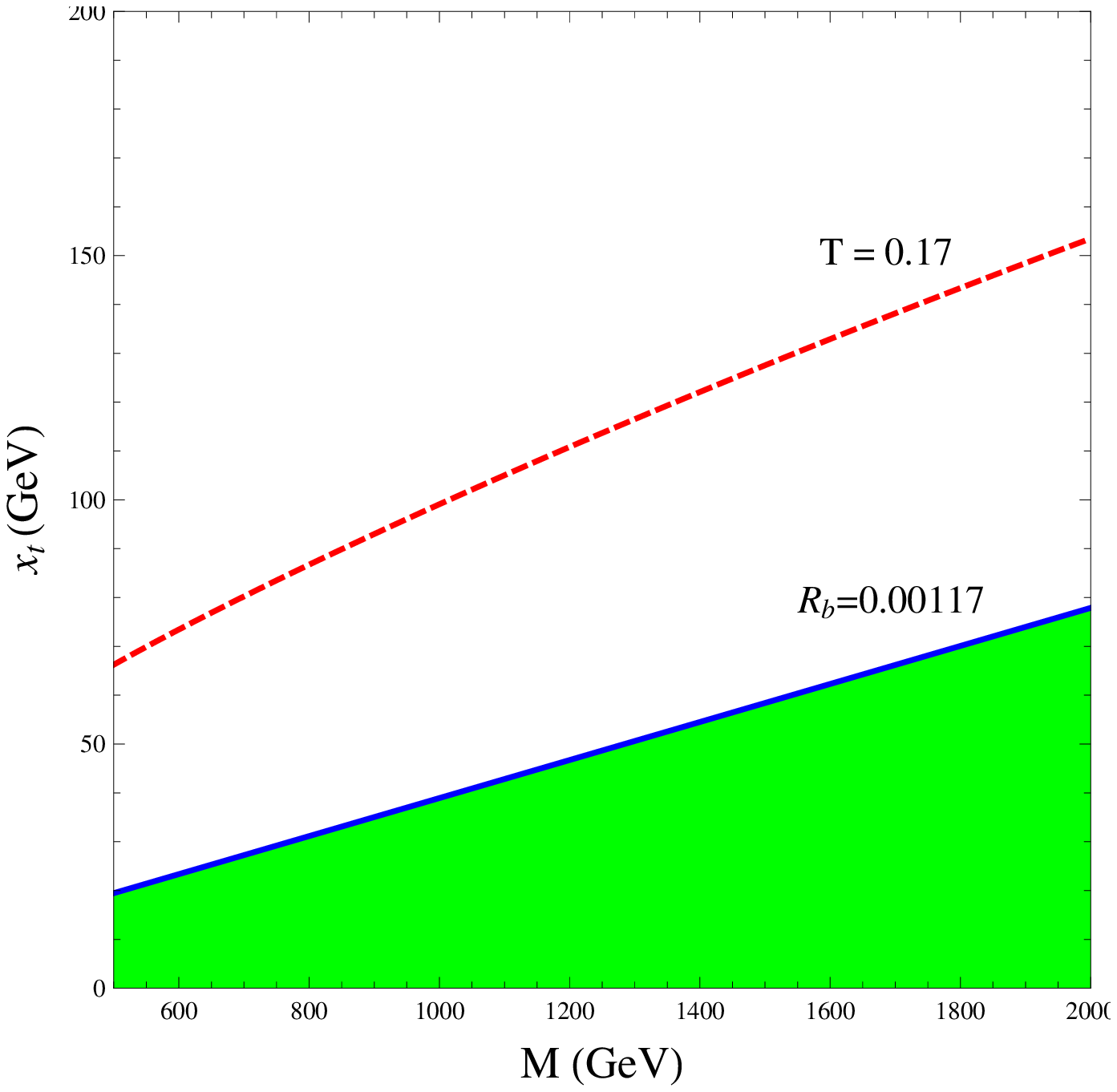}
\end{center}
\caption{$T $ parameter and $R_b$  constraints  for $M$  and $x_t$ in the triplet $\mathcal{T}_X$  vector-quark model . \label{TRb}}
\end{figure}
Another source of $T$ parameter contribution comes from the Higgs mass deviation from the its reference value  $ m_{href} = 120 ~\mbox{GeV} $:
\beq
\Delta {T_h} =  - \frac{3}{{16 \pi c^2}}\log \left( \frac{m_h^2}{m_{href}^2} \right) ~. \label{TH}
\eeq
The electroweak precision measurement gives  the fit  for the $T$
parameter, $ T =0.05 \pm 0.11 ~$~\cite{EW}. Assuming the Higgs
mass is $125~ \mbox{GeV} $ \cite{Ahiggs,Chiggs} , and
adding two sources of  $T$ parameter contribution, the contours of
$T$-parameter  in $ (M, x_t) $  parameter space for the singlet
vector-quark and doublet vector-quarks are plotted in
Figure \ref{TPara}(a-d).  For the singlet scenario as depicted by
Fig.~\ref{TPara}(a),  only the Higgs gives a very small
negative contribution and the singlet top quark  can achieve
larger positive contributions,  therefore we get a  positive $T$-parameter
deviation in the interested vector-quark mass region.  Requiring $ T < 0.17 $ and varying the gauge
invariant mass $M$ in the range of $ (500 -  1000 ) ~ \mbox{GeV}
$,  the upper limit for the top quark mixing parameter  $ x_t $
should be in the range of $ (97.4 - 153.4) \mbox{GeV} $.  Since
the mixing pattern and hyper charge  assignment determine the
$SU(2)$ gauge bosons currents, we see another situation  in the
$\mathcal{D}_X$ doublet vector quark scenario.  The Fig.~\ref{TPara}(b) illustrates the $T$ parameter constraint for the $\mathcal{D}_X$ doublet model with one  mixing  parameter
$x_t $.  The extra $5/3 $ charged heavy quark $X$ does not mix with any SM quarks,  but they will contribute to
 $\Delta T_{\mathcal{D}_X} $.  The heavy quark loop effects
achieve notable positive $T$ parameter deviation when $x_t$ is
large while they give notable negative  $T$ parameter deviation
when $x_t$ is small.  Requiring $ - 0.07 < \Delta T < 0.17 $,
two stringent  green bands in the  $(M, x_t ) $  plane are
permitted.  We are more interested in the lower
green band located in the small $x_t $ region.   As the gauge
invariant mass varies in the range of $ (500 -  1000) \mbox{GeV}
$,  the negative $T$ parameter  bound  constrains $x_t $  to be
in  the range of  $ (54 - 78) \mbox{GeV} $.  For the doublet model with one heavy top quark and one
heavy bottom quark, there will be two independent mixing
parameters  $x_t $ and $x_b$ for the up type quark and down type
quark individually.  $ T $ parameter mainly constrains 
$x_t $  while  $x_b$  is constrained by 
$z \to  b \bar  b $  which will be considered following.  In the
Fig.~\ref{TPara}(c),  the cyan curve, green curve, red curve
and the blue curve depict the  upper bound $T =0.17 $  in the $(M,
x_b)$ parameter space for four sample values $x_t = 80 , 90 , 100,
120 ~ \mbox{GeV }$.  As we can see, in the low $x_b$ region allowed by $R_b$ constraint, $T$ parameter is not
sensitive to  $x_b $  but sensitive to $x_t $ and the upper bound for  $M
$ will increase as we enhance the value of   $x_t $ .  Fig.~\ref{TPara}(d) plots the $\Delta T_{\mathcal{D}_2}  $  dependence on $M$ and $x_t$ with the fixed value $x_b = 20
~\mbox{GeV} $.  Similar to the singlet model  the
$T$ parameter mainly  receives positive contribution from the
heavy fermions in the $\mathcal{D}_2$ doublet model.  After
imposing the upper bound $ T <  0.17 $,  we can find that $x_t$
is constrained to be in the range of $(81- 123)~\mbox{GeV}$ in
correspondence with $M$  in the range of  $ (500 - 1000
)\mbox{GeV} $.  Finally, we discuss the situation in the $\mathcal{T}_X$ triplet model.  Only
the heavy top  and the heavy  bottom in the triplet model will mix  with the SM chiral quarks 
and two mixing parameters are related to each other by virtue of $x_b =
\sqrt 2 x_t$.  The $T$ parameter dependence on $M$ and $x_t$ is
plotted in Fig.~\ref{TRb}, which shows that it  almost does not
put any  constraint in the parameter space compared with the tight
$R_b$ requirement.

In the situation with one pair of heavy bottom quark present, we need to consider the constraint from the $Z \to b \bar b $.   Both the corrections to  $Z$-$b_L$-$b_L$  couplings and the corrections to $Z$-$b_R$-$b_R$ couplings in the doublet model as well as  in the triplet model  are induced through the Yukawa mixing between the vector bottom quarks and the chiral bottom quarks.  The $ \delta g_L^{NP} $ and  $\delta g_R^{NP}$ can be translated to be  the deviation  $\delta R_b$  by the following formula Eq.~[\ref{Rb}]:
\bea
\delta {R_b} &=& 2{R_b}(1 - {R_b}) \left (\frac{{{g_L}}}{{g_L^2 + g_R^2}}\delta g_L^{NP} + \frac{{{g_R}}}{{g_L^2 + g_R^2}}\delta g_R^{NP}\right)   \label{Rb} \\ \nonumber \\
 g_L &=&  \frac{g}{c_W} (-\frac{1}{2} + \frac{1}{3} s_W^2 )   \qquad  \qquad  g_R = \frac{1}{3}  \frac{g}{c_W}  s_W^2 ,    \eea
where $g_L $ and $g_R$ are  $Z $ gauge bosons couplings to the left handed bottom quark and the right handed bottom quark in the Standard Model.  $R_b$ is defined as $\Gamma(Z  \to b\bar b) / \Gamma(Z \to \mbox{hadrons}) $  with  its  SM value  $R_b = 0.21578 ^{+ 0.0005}_{-0.0008}$.  The deviation $\delta R_b$  due to the new physics effects is constrained by electroweak measurements to be \cite{EW} :
\bea
\delta {R_b} &=& 0.00051 \pm 0.00066  \eea
We can write down  the  $Z$-$b_L$-$b_L$ couplings and  the $Z$-$b_R$-$b_R$ couplings  in the $\mathcal{D}_2$ doublet model and in the $\mathcal{T}_X$ triplet model. For the doublet model we have:
\bea g_{ZbL} & =&  \frac{g}{c_W} (- \frac{1}{2}  + \frac{1}{3} s_W^2 ) = g_{ZbL}^{sm}   \label{DbL}\\
 g_{ZbR}   &=&    -\frac{1}{2 } \frac{g}{c_W}  \sin^2 \theta_d^R +  \frac{1}{3} \frac{g}{c_W}  s_W^2  =  -\frac{1}{2 } \frac{g}{c_W}    \sin^2 \theta_d^R + g_{ZbR}^{sm}   \label{DbR}
 \eea
while for  the triplet model  we have  :
\bea
g_{ZbL}  &=&  -\frac{1}{2 } \frac{g}{c_W}    \sin^2 \theta_d^L  +  \frac{g}{c_W} (- \frac{1}{2}  + \frac{1}{3} s_W^2 ) =  -\frac{1}{2 } \frac{g}{c_W}   \sin^2 \theta_d^L  + g_{ZbL}^{sm}  \label{TbL}  \\
 g_{ZbR}  &=&    -  \frac{g}{c_W}  \sin^2 \theta_d^R    + \frac{1}{3}  \frac{g}{c_W}  s_W^2  =  -  \frac{g}{c_W}    \sin^2 \theta_d^R+ g_{ZbR}^{sm}  \label{TbR}
 \eea
\begin{figure}
\begin{center}
\includegraphics[width=0.35 \hsize]{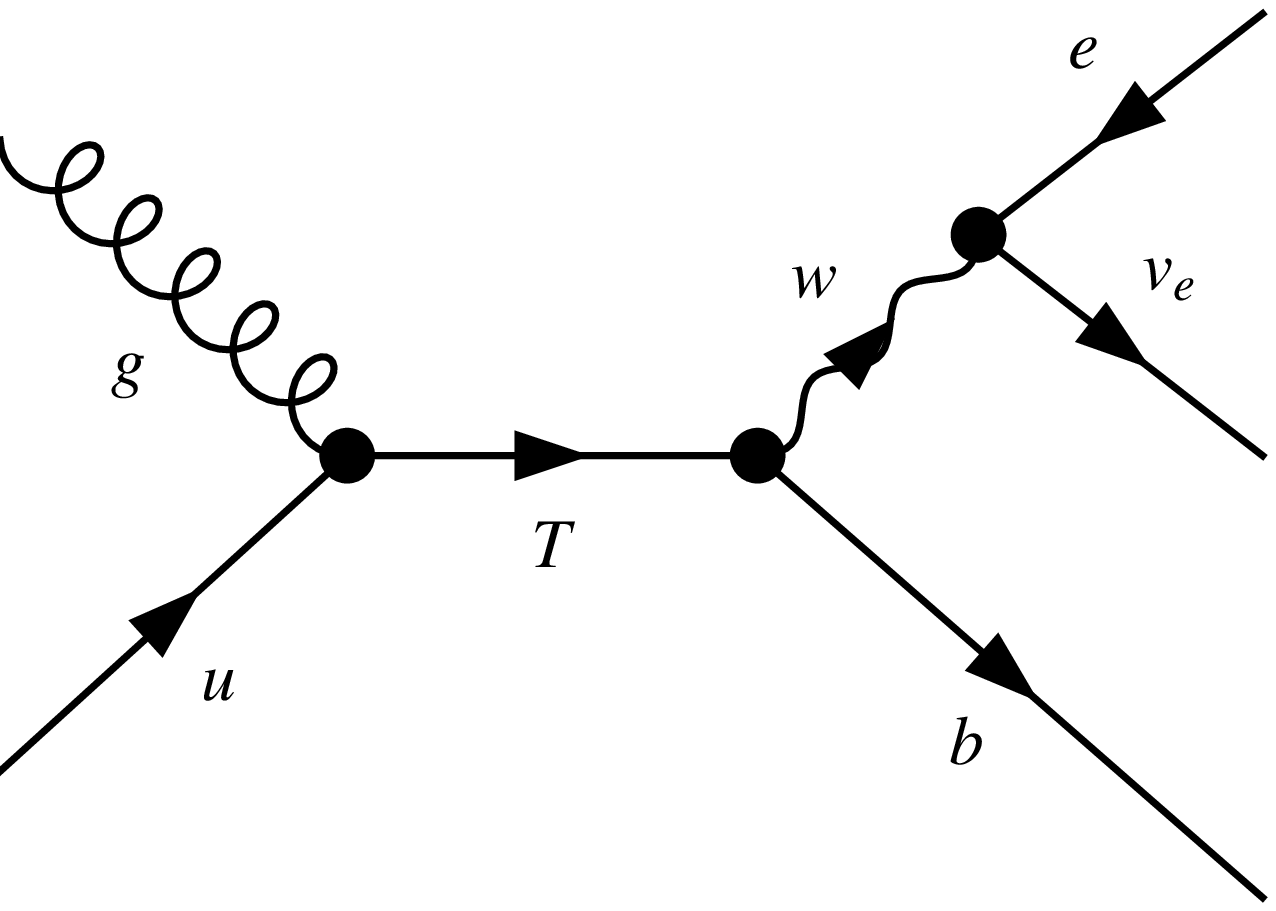}
\end{center}
\caption{Feynman diagram for single-$T$ production and decay ($ u g \to T \to  b W^+ \to b \ell^+ \nu_\ell $). \label{ugbw}}
\end{figure}
Substituting Eq.[\ref{DbL}- \ref{TbR}] into Eq.[\ref{Rb}], we  get the $R_b$ deviation due to the vector-quarks in the doublet model  and  in the triplet model respectively, which in turn  gives stringent constraints for $x_b$ and $x_t $ in each scenario.  As we can see from Fig.~\ref{TPara}(c) which depicts the situation for the $\mathcal{D}_2$ doublet model,  the negative $R_b$ bound limits the $x_b$ parameter to be in the range of $(23 - 46) ~\mbox{GeV}  $ as  $M $ varies from $500 ~\mbox{GeV} $ to $ 1000 ~\mbox{GeV} $.   Fig.~\ref{TRb} shows that  it is  the upper limit of  $R_b < 0.00117$  which puts  constraint on $ M $ and $ x_t $  for the $\mathcal{T}_X$ triplet model. The green band is the allowed parameter space, i.e.  for $ 500 ~\mbox{GeV} < M < 1000 ~\mbox{GeV} $, we need $ 20 ~\mbox{GeV} < x_t < 39 ~\mbox{GeV} $.

\begin{table}
\begin{center}
\caption{Cross section (fb) of the signal process $ pp  \to T \to
b  W^+  \to b \ell^+ \nu $ and SM backgrounds for various
kinematics cuts. The top table is for the singlet vector-quark
with the choice of $ f_L =0 $, $ f_R=0.2 $ and $ m_T =
800~\mbox{GeV} $. The middle table is for the doublet vector-quark
with the choice of $ f_L =0.3  $, $ f_R=0 $ and $ m_T =
800~\mbox{GeV} $ and with an addition input $x_b =20~\mbox{GeV}$.
The bottom table is for the SM backgrounds.} \vspace*{0.5cm}
\begin{tabular}{ccccccc} \hline
$ x_{t} ~(\mbox{GeV}) $ & no cuts  &  basic  cuts & $ P^b_T> 200  ~\mbox{GeV} $ & $ \Delta R_{\nu l}  <1. 5 $    & $  \Delta m_W < 10 \mbox{GeV}  $
\\ \hline
40 \mbox{GeV} &  211.8 & 73.643 & 69.841 & 66.283  & 66.283 \\ \hline
60 \mbox{GeV} &  223.1 & 77.059 & 73.121 & 69.596  & 69.596 \\  \hline
80 \mbox{GeV} &  228.02 & 80.046 & 76.056 & 72.237  & 72.237 \\  \hline
100 \mbox{GeV} &  230.44 & 80.216 & 76.518 & 73.199  & 73.199 \\  \hline
120 \mbox{GeV} &  232.18 & 81.890 & 77.595 & 74.019  & 74.019  \\ \hline
\end{tabular}

\vspace*{0.5 cm }

\begin{tabular}{ccccccc} \hline
$ x_{t} ~(\mbox{GeV}) $ & no cuts  &  basic  cuts & $ P^b_T > 200  ~\mbox{GeV} $ & $ \Delta R_{\nu l}  <1. 5 $   & $  \Delta m_W < 10 \mbox{GeV}  $
\\ \hline
40 ~\mbox{GeV} & 166.25 & 57.847 & 54.921 & 52.336  & 52.336 \\ \hline
60 ~\mbox{GeV} & 114.36 & 39.694 & 38.088 & 36.115  & 36.115 \\  \hline
80 ~\mbox{GeV} & 88.792 & 30.314 & 28.968 & 27.552  & 27.552 \\ \hline
100 ~\mbox{GeV} & 75.728 & 25.369 & 24.426 & 23.211  & 23.211 \\ \hline
120 ~\mbox{GeV} & 68.41 & 22.599 & 21.70  & 20.732  & 20.732  \\ \hline
\end{tabular}

\vspace*{0.5 cm }

\begin{tabular}{cccccc} \hline
 SM bg  & no cuts  &  basic  cuts & $ P^b_T > 200  ~\mbox{GeV} $ & $ \Delta R_{\nu l}  < 1.5 $   & $ \Delta m_W < 10 \mbox{GeV}   $
\\ \hline
$ w^+j  $ & $ 1.95*10^{7}  $  & $ 5889.06 $  & $ 175.97 $  & $ 175.97 $  & $ 175.97$
\\  \hline
$ w^+b j   $ & $ 2774.2  $  & $ 80.32  $  & $ 1.942 $  & $ 1.248 $ & $ 1.248 $
\\  \hline
$ t  j   $ & $ 29619  $  & $ 5386.22  $  & $ 42.947 $  & $ 7.405 $ & $ 7.405 $
\\  \hline
$ t \bar{b}    $ & $ 1025.1  $  & $ 22.14 $  & $ 0.28 $  & $ 0.22 $  & $ 0.22 $
\\  \hline
total  & $ 1.95334 * 10^7  $  & $ 11377.7 $  & $ 221.139 $  & $ 184.843 $  & $ 184.843 $
\\  \hline
\end{tabular}
\end{center}
\label{tab_bw}
\end{table}
In this section we still consider the $T$-quark is produced via
anomalous $q$-$T$-$g$ couplings, but it could decay through
the Yukawa interaction shown above. Note that after the mixing the
$T$-quark as an $s$-channel resonance is no longer purely
polarized. After adding the Yukawa mixing  the $T$-quark gets
 three electroweak decay modes opened: $t h$,  $ t Z$ and $ b W^+
$.  Here we proceed to analyze the process of $ g q \to T \to b
W^+ $ with a subsequent $W$-boson leptonic decay (see Fig.
~\ref{ugbw} for Feynman diagrams), yielding a collider signature of
$ b \ell^+ \met $. The main SM backgrounds are (1) $ W^+ j $ where
the light jet can mimic bottom quark; (2) direct $ W^+ b j $  production as well as
$t$-channel single-top production $qb\to q^\prime t\to W^+bj$,
in both cases we require that  one of  non-b tagged jets  escapes
detecting; (3)  $s$-channel single-top production channel
$q\bar{q}^\prime \to t\bar{b}\to W^+ b\bar{b}$, where one of the
bottom quarks escapes the detection. The basic cuts to trigger the
events are:
\bea
&& p_T^b > 50~{\rm GeV},\qquad \left| {{\eta _b}} \right| < 2.0,  \nonumber \\
&& p_T^l > 20~{\rm GeV},\qquad \left| {{\eta _l}} \right| < 2.4,  \\
&& \Delta {R_{bl}} > 0.7 . \nonumber
\eea
The cut table for the signal event (singlet case  and doublet case) at the bench mark point $m_T = 800~{\rm GeV}$ as well as  for the background is shown in Table~\ref{tab_bw}. The third column shows the numbers of event after the basic cuts. We demand  $b$-tagged when imposing the basic cuts where the veto cuts for the non-intrinsic background events are also included.  The $b$-tagging requirement effectively reduces the $Wj$ background by a factor of $0.5\times 10^{-3}$, while it still keeps about  $1/3$ signal events.
\begin{figure}[tb]
\begin{center}
\includegraphics[width=0.45 \hsize]{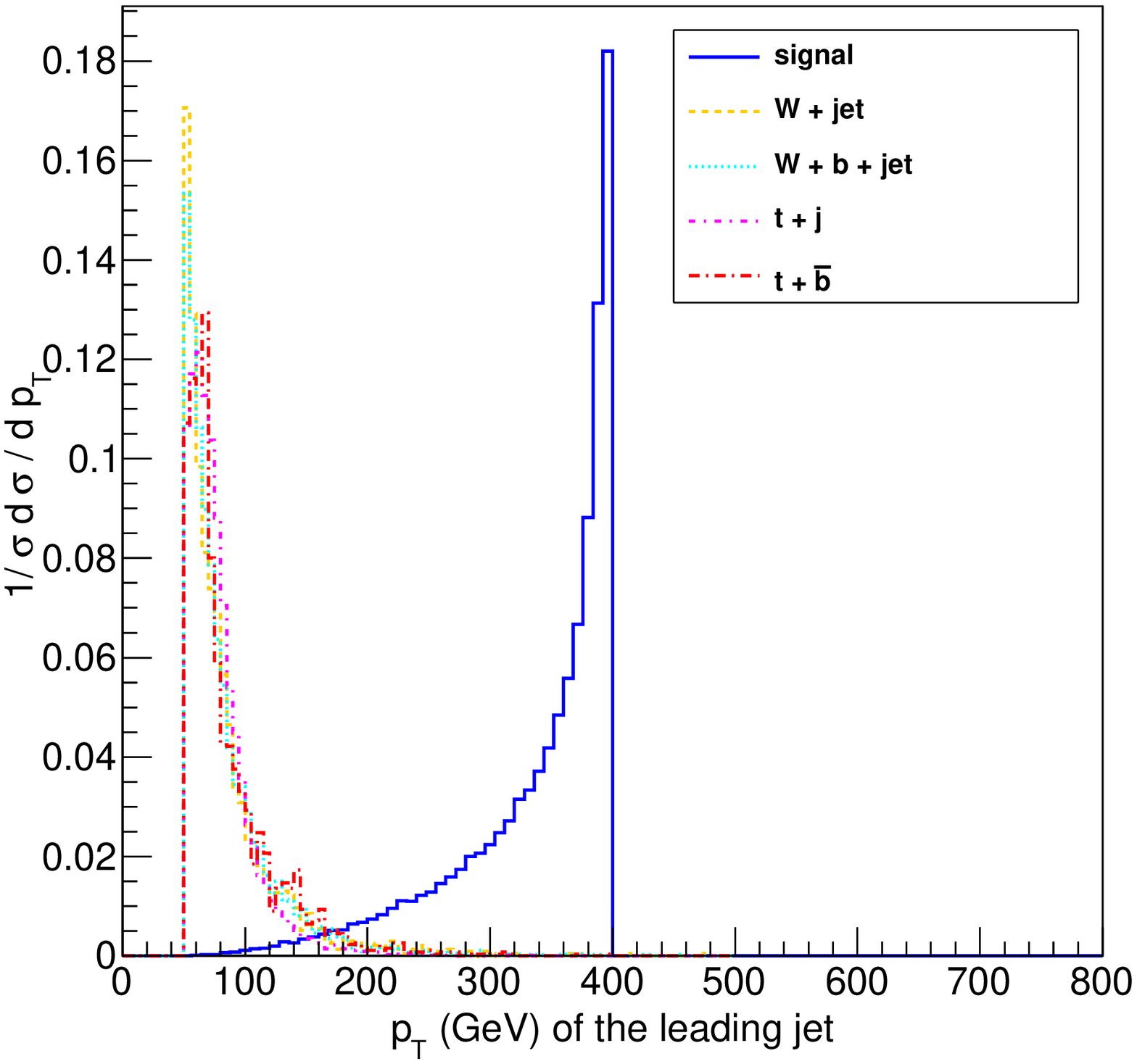}
\includegraphics[width=0.45 \hsize]{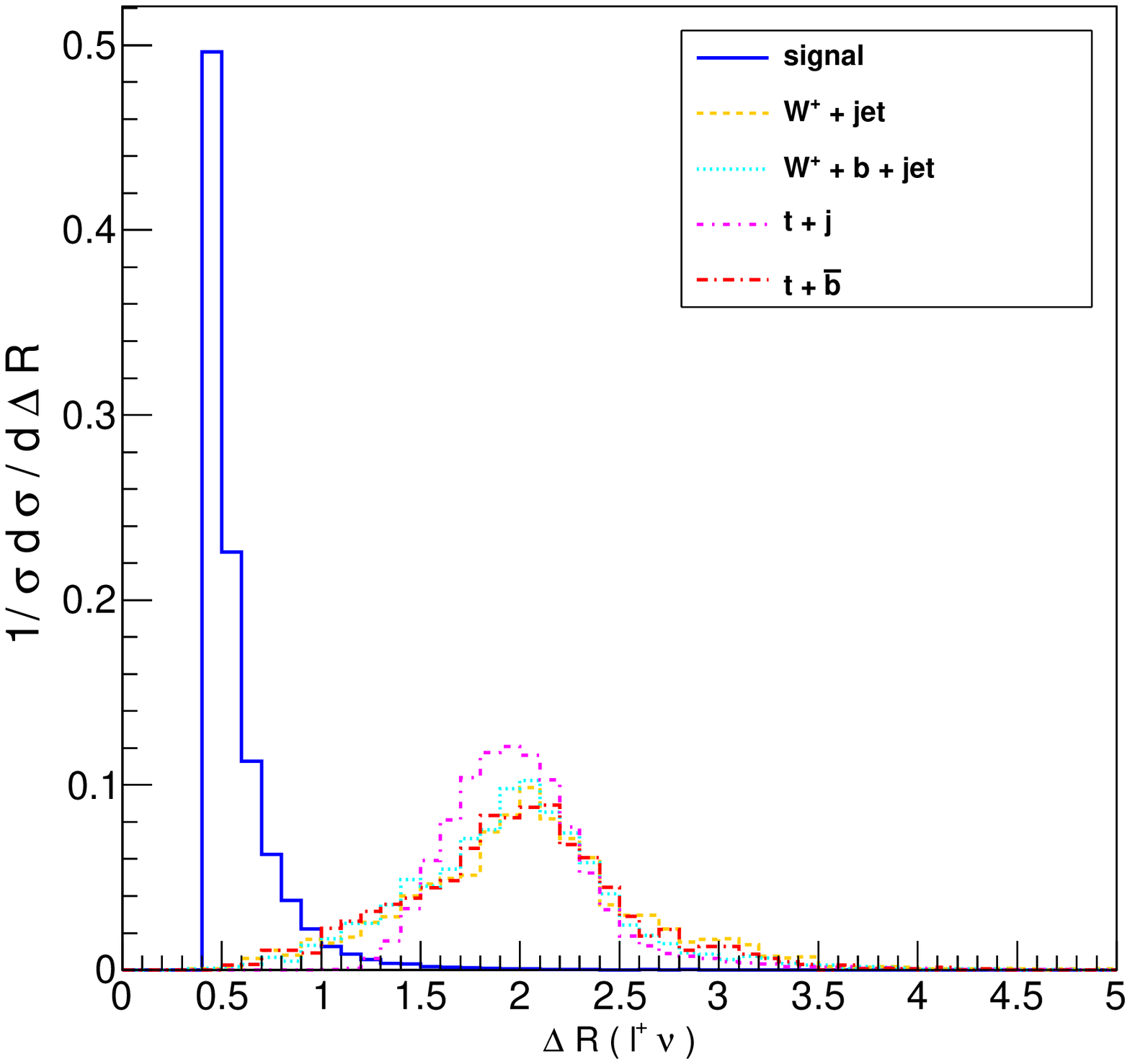}
\end{center}
\caption{ (a) The $p_T$ distribution of the leading jet  after the basic and veto selection cuts;  (b) the distribution of the $\Delta R ( \ell^+  \nu ) $ between the lepton and the reconstructed neutrino after the basic and veto selection cuts. In both plots we choose $m_T = 800 ~\mbox{GeV} $ and each distribution is normalized.
\label{ptdr}}
\end{figure}

In the signal event the $b$-jet comes from the heavy vector-quark decay. It carries a large $p_T$. On the contrary, the $b$-jets in the background events, either from the gluon splitting or from the top-quark decay, exhibit a relatively soft $p_T$. The difference can be seen  from Fig.~\ref{ptdr}(a) where the $p_T$ distribution of the $b$-jet is plotted.  Following the basic cut and $b$-tagging,  we impose a hard $p_T $ cut on the $b$-jet:
\bea
p_T (b) > 200~{\rm GeV} .
\eea
As shown in Table~\ref{tab_bw} this cut reduces the background  more than one order of magnitude, but it leaves the signal events almost untouched.   The reason that we do not consider the non-intrinsic background events from $ W^+ b \bar b $ and $W^+ Z $ is that $ b \bar b $ from either gluon splitting or $ Z$ bosons decay can not simultaneously  pass the veto cuts as well as the hard $p_T (b) $ cut .

To fully reconstruct the signal events, the kinematics of the invisible neutrino is to be determined from the $W$-boson on-shell condition. We pick up the solution with a smaller magnitude in the two-fold solutions. The $W$-bosons from the heavy vector-quark $T$ decay is highly boosted such that its decay products of the lepton and neutrino tend to collimate and yield a small $\Delta R (\ell^+,\nu)$ separation as we can see from the blue-solid curve in Fig~\ref{ptdr}(b).  On the other hand,  the background events are more evenly distributed in the relative large $\Delta R (\ell^+,\nu)$ region.

\begin{figure}[tb]
\begin{center}
\includegraphics[width=0.36 \hsize]{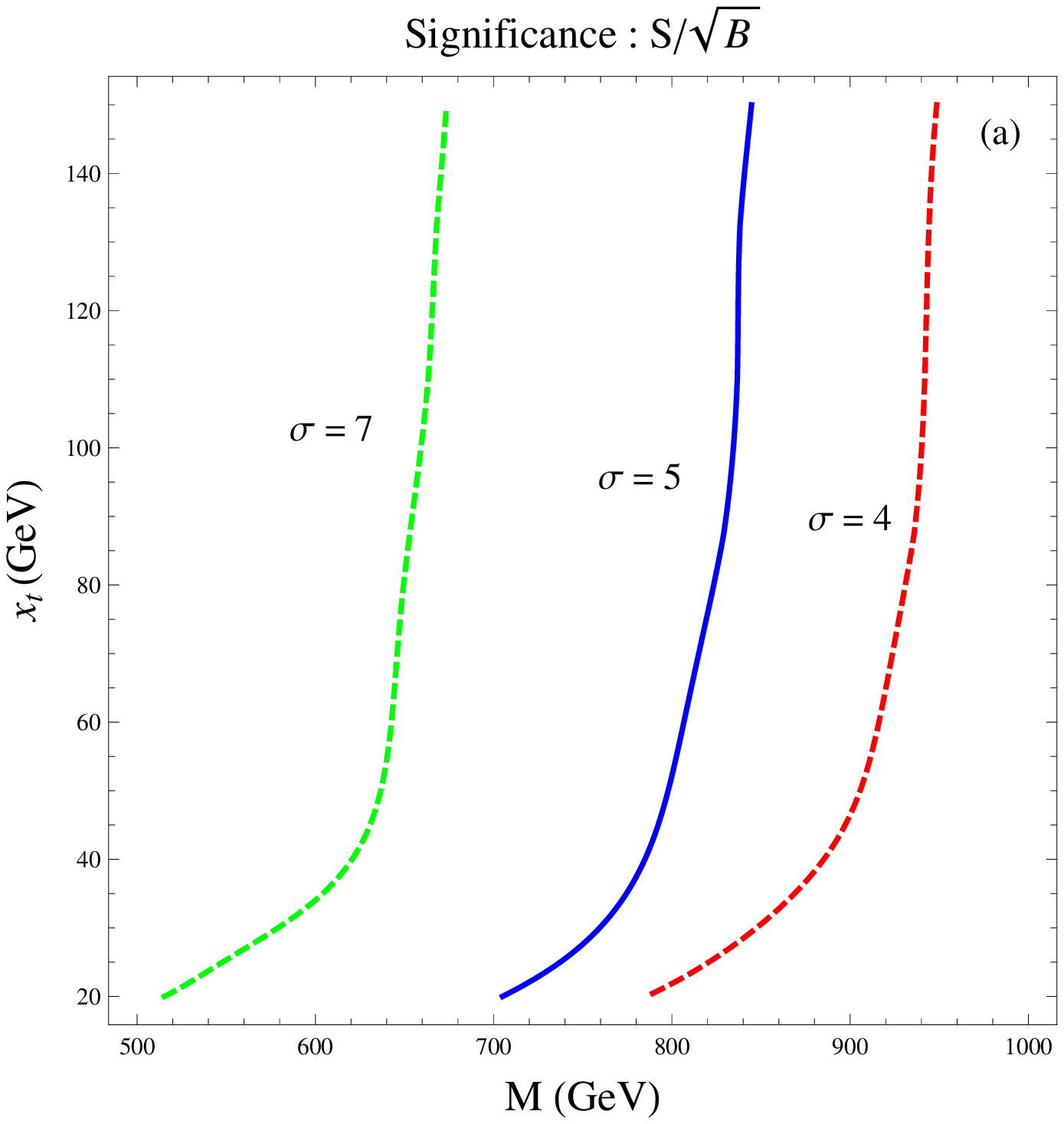}  \hspace{0.5 cm }
\includegraphics[width=0.36 \hsize]{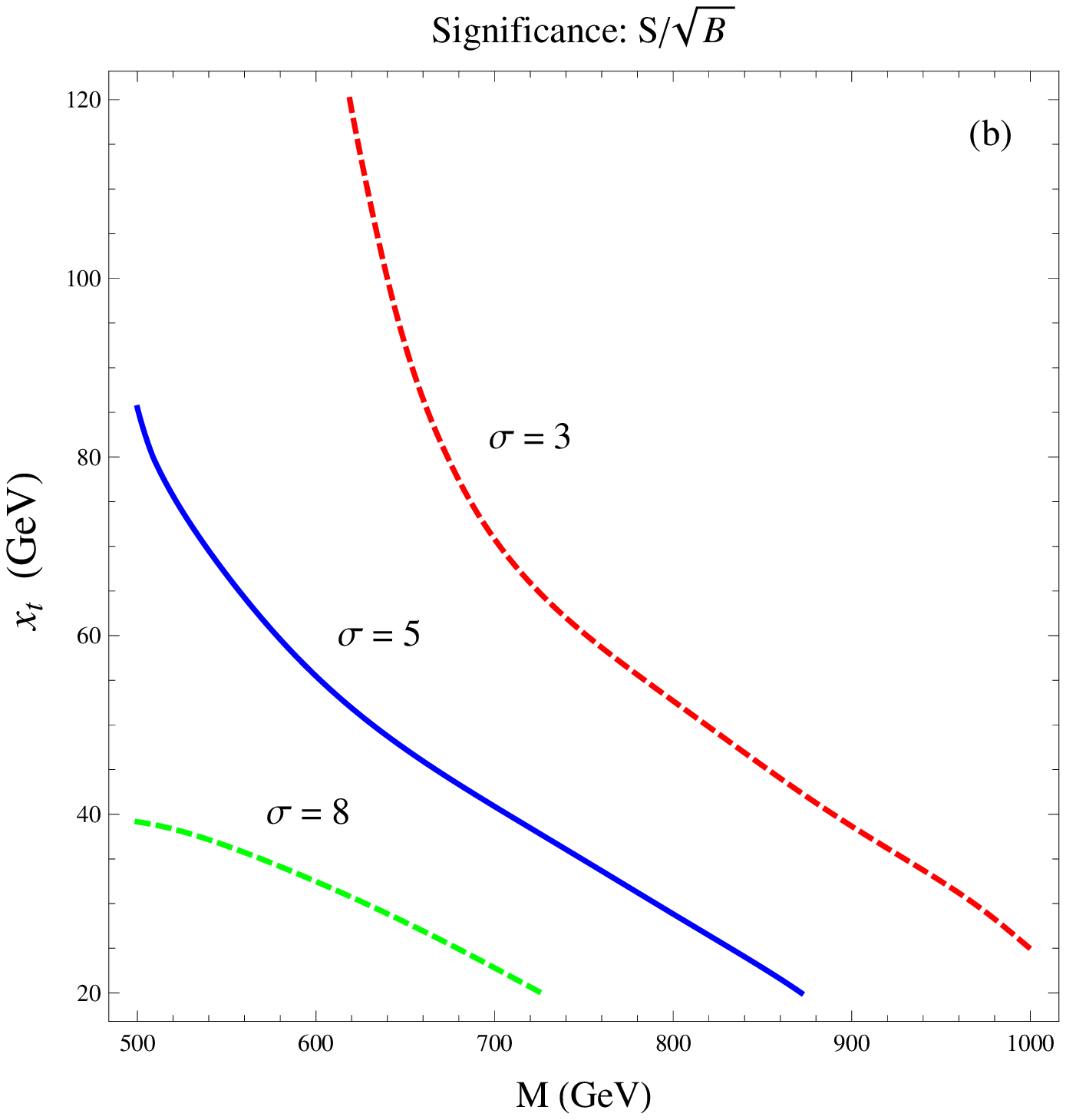}
\includegraphics[width=0.36 \hsize]{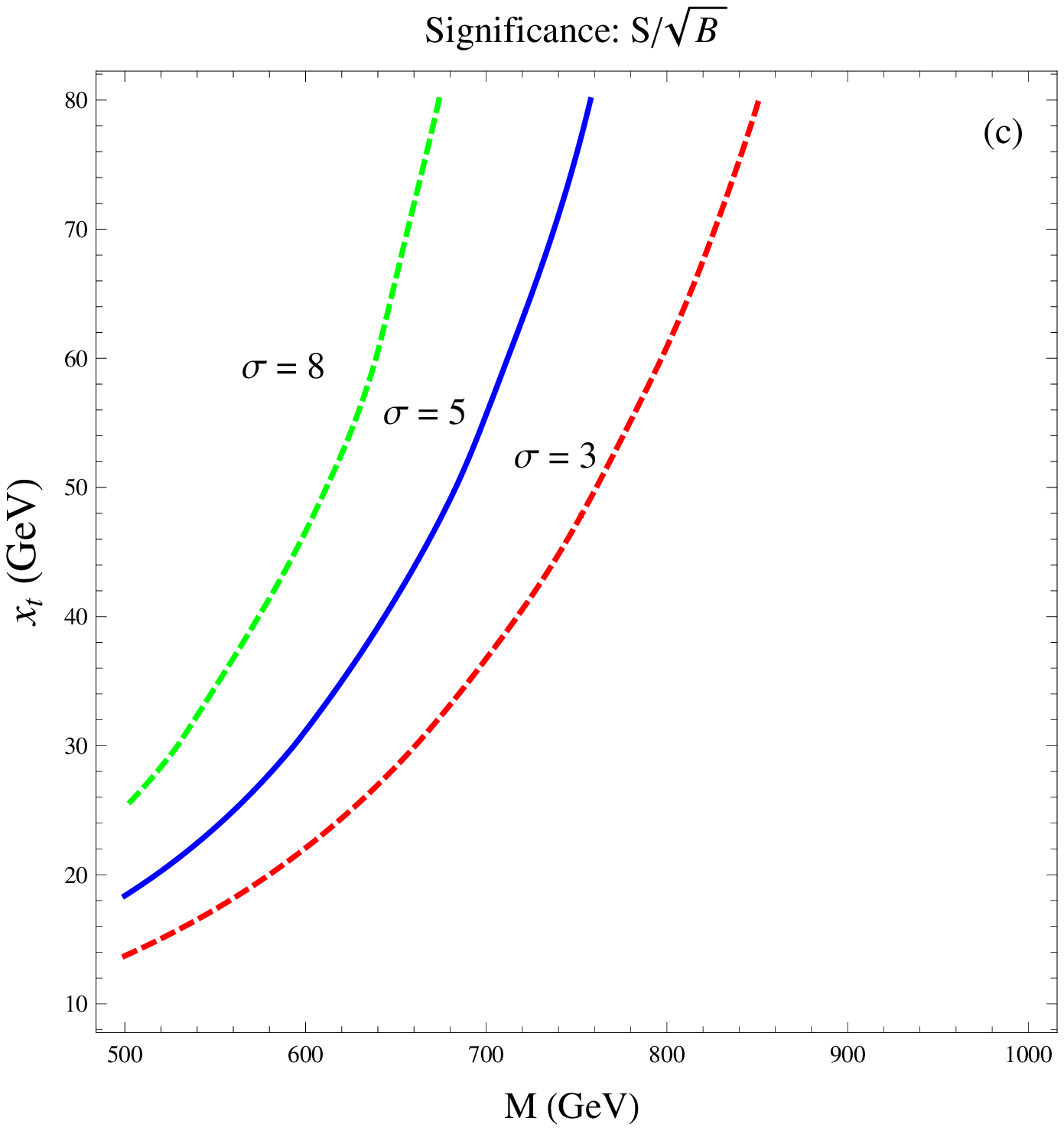}   \hspace{0.5 cm }
\includegraphics[width=0.36 \hsize]{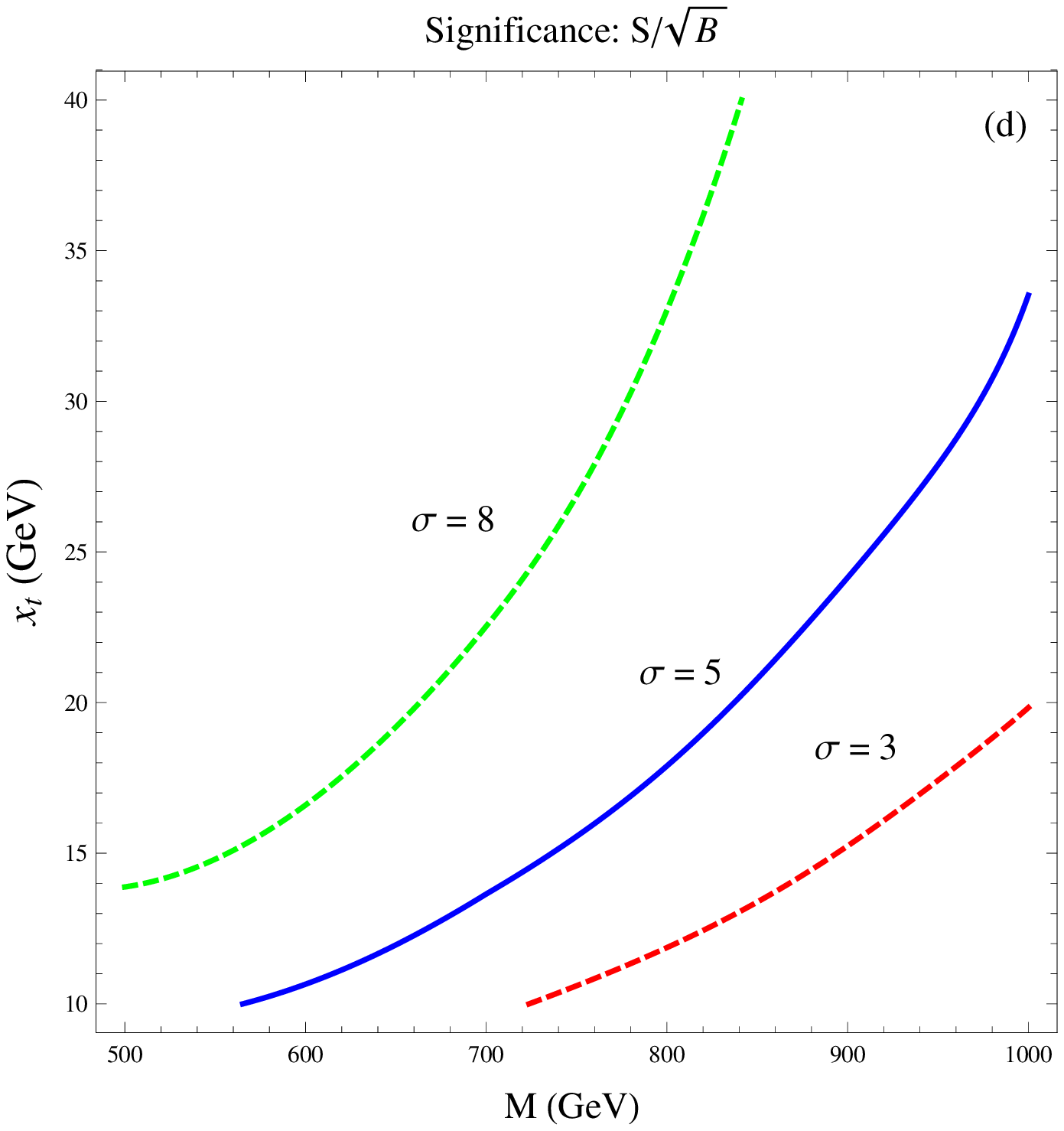}\end{center}
\caption{Significance contour in the $(x, M)$ parameter space: (a) the singlet $\mathcal{U}_1$ model ($f_R =0.2 $); (b) the doublet $\mathcal{D}_2$ model ($f_L =0.3 $ and $x_b = 20 ~\mbox{GeV} $); (c) the doublet $\mathcal{D}_X$ model ($f_L =0. 6 $) ;  (d) the triplet $\mathcal{T}_X$ model ($f_R =0.3 $).
\label{SGbw}}
\end{figure}
Fig.~\ref{SGbw} displays the significance contours at the 14~TeV
LHC for several vector-quark models. The doublet $\mathcal{D}_Y$
is  ignored since there is no heavy $T$-quark in  that model,  and
the  triplet $\mathcal{T}_Y$ models  is  not considered  either
because it has  a very small branching ratio for the decay of $T
\to W^+b$.  We will emphasize some characteristics  in
those contours and  illustrate the reasons as
follows. The $\mathcal{U}_1$, $\mathcal{D}_X$ and $\mathcal{T}_X$
models exhibit a similar pattern of the discovery potential, i.e.
$x_t$ increases with $M$ for a fixed significance. While the
doublet $\mathcal{D}_2$ model shows a different pattern as $x_t$
decreases with $M$. The difference is caused by the decay
branching ratio of the vector-quark in various
models~\cite{Cacciapaglia}. The branching ratio of  $T \to W^+ b$
decay always increases as the value of $x_t$ is enhanced in the
$\mathcal{U}_1$, $\mathcal{D}_X$ and $\mathcal{T}_X$ models.
However in the $\mathcal{D}_2$ model the corresponding branching
ratio can be approximated by  $  x_b^2 / (x_b^2 + 2 x_t^2) $ in
the large $M$ limit.  There is a tension exists between $x_b$ and
$x_t$ and the branching ratio of $T\to W^+b$ decreases with $x_t$
for a fixed  $x_b$ value,  which leads to the slopping pattern
displayed in Fig.~\ref{SGbw}(b).   As we can see from the
significance plot for the $\mathcal{U}_1$ singlet model with $f_R
=0.2 $,  that  the significance contour is not sensitive to the
mixing  parameter in the large $x_t $ region, such that the $\sigma =5 $
discovery limit  curve is approaching to be vertical around  $M= 800 ~ \mbox{GeV} $. Note that the
$\mathcal{D}_X$ doublet model has a smaller discovery potential as
indicated by its magnetic coupling.  The reason is that  the
branching ratio of $T \to W^+ b $ is  suppressed by a factor of
$m_t^2 / M^2 $  in that model  as compared with the other three ones.

\begin{figure}[htb]
\begin{center}
 \includegraphics[width=0.45 \hsize]{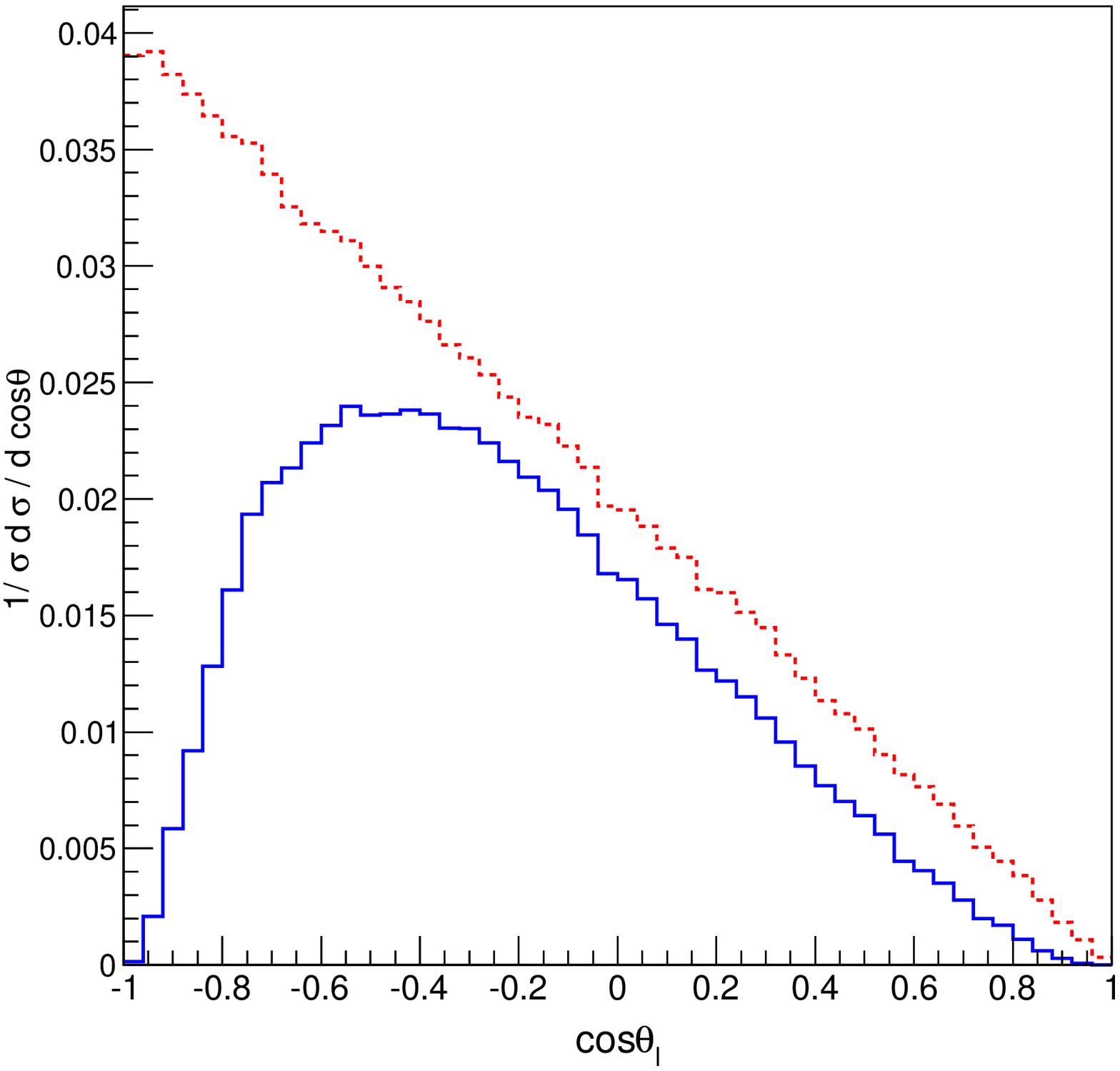}
  \includegraphics[width=0.45 \hsize]{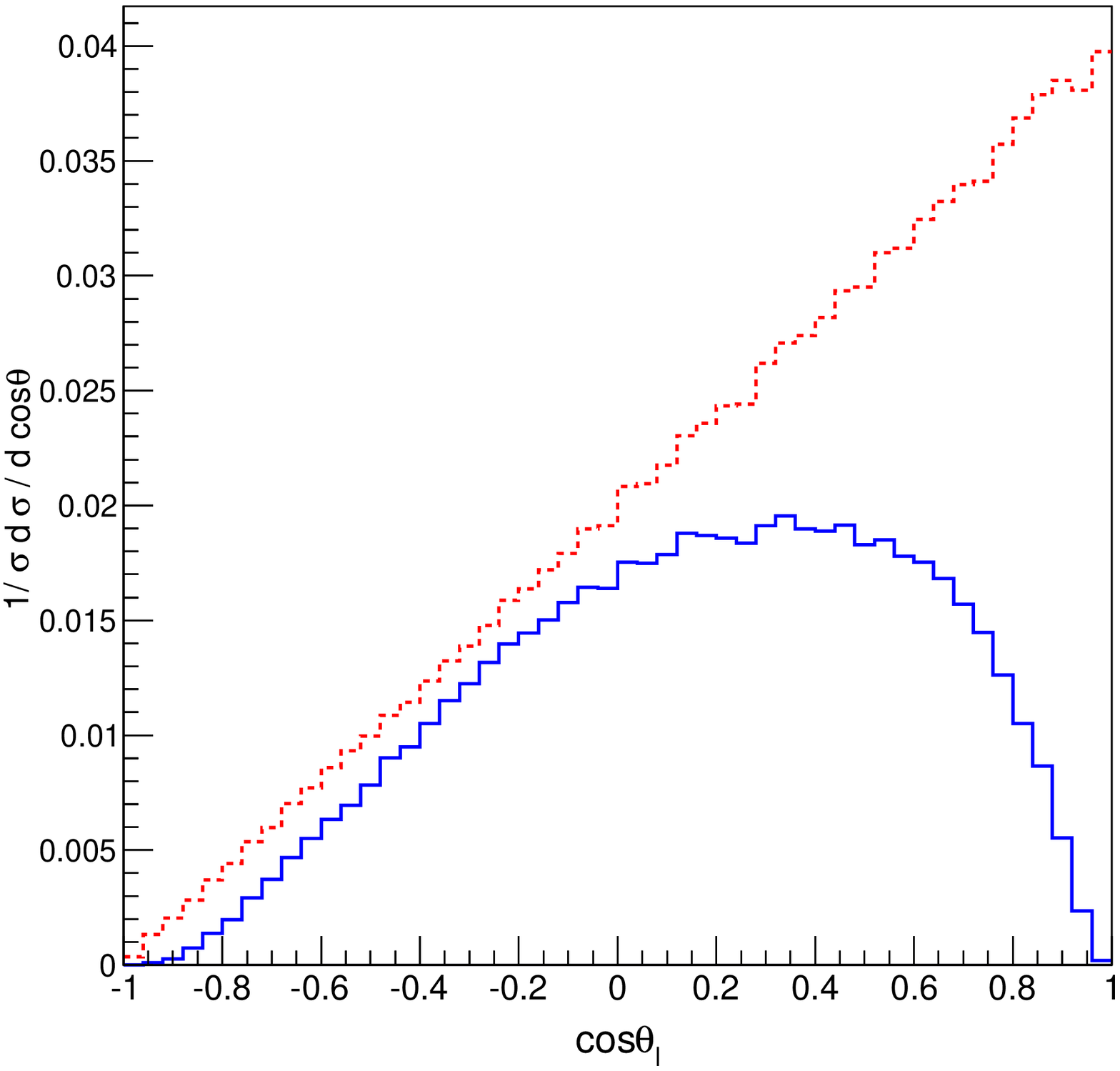}
   \end{center}
\caption{The $\cos \theta_{\ell}$ distribution between the lepton and gluon moving directions  with no cut (dashed line)  and after the mass-window cut (solid line): (a)  the singlet $T$-quark, (b) the doublet $T$-quark (when there is  no heavy bottom quark or its effect can be ignored). }
\label{costlbw}
\end{figure}
It is convenient  to analyze  the charged-lepton angular distribution  between the lepton and the gluon moving directions in the center of mass frame,  which  can be used to distinguish the chiral property  of   $T$-$b$-$W$ couplings.  The differential cross sections with respect to  $\cos \theta_\ell $  for the $u g \to b \ell^+ \nu $  in  vector-quark models  in the limit of $m_W \ll  \sqrt s $  are found to be:
 \bea
\frac{1}{\hat \sigma}\frac{ d \hat \sigma(u_R g \to b\ell^+ \nu) }{d \cos \theta_\ell } &=&  \frac{g_{W}^{L ~2}}{g_{W}^{L ~2}+ g_{W}^{R ~2}} (1- \cos\theta_\ell)  +  \frac{g_{W}^{R ~2} }{g_{W}^{L ~2}+ g_{W}^{R ~2}} (1+\cos \theta_\ell ) \nonumber  \\  &-&  \frac{g_{W}^{R ~2} }{g_{W}^{L ~2}+ g_{W}^{R ~2}} \cdot  \mathcal{O} (m_W^2/ s ) \cdot  \cos \theta_\ell  \label{Rtheta}
\\ \nonumber \\
\frac{1}{\hat \sigma} \frac{ d \hat \sigma(u_L g \to b\ell^+ \nu) }{d \cos \theta_\ell } &=& \frac{g_{W}^{L ~2}}{g_{W}^{L ~2}+ g_{W}^{R ~2}} (1+ \cos\theta_\ell)  +   \frac{g_{W}^{R ~2} }{g_{W}^{L ~2}+ g_{W}^{R ~2}} (1 -\cos \theta_\ell)  \nonumber  \\  &+&  \frac{g_{W}^{R ~2} }{g_{W}^{L ~2}+ g_{W}^{R ~2}} \cdot  \mathcal{O} (m_W^2/ s ) \cdot  \cos \theta_\ell \label{Ltheta}
 \eea
Note that the $\mathcal{O} (m_W^2/ s )$ correction is only for the term proportional to the right-handed $T$-$b$-$W$ coupling. The detail derivation of exact result for the leptonic angular distribution in the rest frame of the heavy top quark  is put  in  the Appendix.  Fig.~\ref{costlbw} displays the $\cos\theta_{\ell} $ distributions  with no cut  and  after the mass-window cut.   For the singlet $\mathcal{U}_1$ model, since the anomalous $u$-$g$-$T$ interaction is right handed and  the coupling of $T$-$b$-$W$ is purely left handed, the angular distribution of $\cos\theta_{\ell}$ should be proportional to $(1-\cos \theta )/2 $ as indicated by Eq. [\ref{Rtheta}] ; On the other hand, the anomalous $u$-$g$-$T$ interaction is left handed  and the coupling of $T$-$b$-$W$ is also purely left handed in the doublet $\mathcal{D}_X$ model,  therefore the lepton angular distribution should be proportional to   $(1+\cos \theta)/2 $,  consistent  with the analytic result in Eq. [\ref{Ltheta}]. The situation is more complicated in  the doublet $\mathcal{D}_2$ model. In that model  we have $g_W^L / g_W^R \approx ( x_t m_t ) / (x_b  M) $  such that the angular distribution shows the superposition of $(1 \mp \cos \theta )/2 $. In one limit  $ x_b\gg x_t $,  the coupling of $T$-$b$-$W$ is mainly  right-handed and the angular distribution is similar to the singlet $\mathcal{U}_1$.  In the other limit $x_t \gg x_b $, the decay of $T\to bW^+$ is possible to be dominated by the left-handed coupling as long as $M $ is not too large, so that  its angular distribution is similar to the doublet $\mathcal{D}_X$.   The leptonic angular distribution should serve as  an effective analyzing power  when the $V-A$ chiral structure of the $T$-$b$-$W$ vertex is not too much modified.

\section{Conclusion}

In this paper,  the  single heavy top quark production via strong  magnetic interaction is studied and we explore the possibility for vector-quarks to be discovered in  both the anomalous decay and electroweak decay channel. The  leptonic angular distribution is  a  favored  analyzing power for identifying the top quark  polarization and distinguish varieties of vector-quark models.  We use the collider simulation to explain that  the  top polarization in the channel of  $T \to t g$  is determined by  the chirality property of the excited quarks.  The Yukawa mixing does not change the situation since  the mixing only dilutes the corresponding branch ratio in each specific model.  However the $V-A$ chiral  structure of $T$-$b$-$W$  is possible to be modified by those Yukawa mixing . We conduct a detail analysis for the parameter space  constrained by the electroweak precision test in various vector-quark models.  The  channel of $T \to b W^+ $  has very less SM backgrounds,  and  the  leptonic angular distribution in that channel  can at least be utilized to distinguish the $ \mathcal{U}_1$ singlet model from the $\mathcal{D}_X $ doublet model.   Once we found the signals of  heavy  $T$ vector-quark  and measured its chirality, we are capable to  reconstruct its mass as a resonance in a  single production process.   Although we have shown that  the LHC  is sensitive to   the signals of  vector like quarks  in the single production channel,  the cross section of such process depends on the coupling strength of  strong magnetic interactions,  otherwise the pair production of heavy top quarks  should be a better discovery  channel .

\begin{acknowledgments}
I thank Qing-Hong Cao and Jian Wang for collaboration in the
early stage. The work is  supported in part by the postdoc
foundation under the Grant No. 2012M510001.
\end{acknowledgments}

\newpage
\appendix

\section{Leptonic angular distribution in c.m. frame}

We derive  the angular distribution for the moving direction of charged lepton relative to the  gluon moving direction in the center of mass rest frame for the process $ u g  \to T \to b \ell^+ \nu $. The $\hat z$ axis is defined by the moving direction of the charged lepton $\ell^+ $  in the c.m. frame of the initial partons .
Since the production and the decaying  can be factorized using the narrow width approximation,
we further have the mass on shell relation i.e. $ k_W ^2= M_W^2 $, such that  the energy and momentum of the bottom quark
is determined by  energy-momentum conservation  to  be: $E_W= \frac{\sqrt s}{2} (1+\frac{m_W^2}{s} )$ and $p =\frac{\sqrt s}{2} (1- \frac{m_W^2}{s} )$.
The four momentum for the initial partons and the bottom quark, the lepton and the intermediate  $W^+$  gauge bosons  can be expressed as:
\bea
{p_g} &=& (\frac{\sqrt s}{2} , -\frac{\sqrt s}{2} \sin \theta_\ell , 0 ,  \frac{\sqrt s}{2} \cos \theta_\ell )    \\
  {p_u} &=&  (\frac{\sqrt s}{2}, \frac{\sqrt s}{2} \sin \theta_\ell,0 , - \frac{\sqrt s}{2} \cos \theta_\ell)  \\
{k_b} &=& ( p , p \sin \theta_{b \ell} \cos \phi_{b \ell} , p \sin \theta_{b \ell} \sin \phi_{b \ell},  p \cos \theta_{b \ell} ) \\
 {k_W} &=& (E_W , -p \sin \theta_{b \ell} \cos \phi_{b \ell} , -p \sin \theta_{b \ell} \sin \phi_{b \ell},  -p \cos \theta_{b \ell}  )  \\
{k_\ell} &=& (E_\ell , 0 , 0,  E_\ell)
\eea
where $\theta_\ell$   is  polar angle  for  the incoming gluon make with the lepton moving direction ($\hat z$-axis), $\theta_{b \ell } $ is angle between the bottom quark and  the lepton and the  $\phi_{b \ell}$ is the corresponding azimuthal angle.  In the left-handed and right-handed strong magnetic interaction cases, the amplitudes squared are calculated to be:
\bea
|M(u_L g \to be^+v)|^2&=& \frac{64}{\Lambda ^4}  g_2^2 v^2 f_L^2 p_g\cdot p_u \left( g_{W}^{ L~ 2}
   s ~ k_{e} \cdot p_u k_{ve}\cdot k_b +  g_{W}^{ R~ 2} M_T^2 ~ k_{ve} \cdot p_g k_{e}\cdot k_b \right) \nonumber \\ && \cdot
 \frac{\pi   \delta ( k_W^2 -m_W^2 )}{\Gamma_W m_W}\frac{C_F \cdot N_c}{((s-M_T^2)^2 + \Gamma_T^2 M_T^2) }
\eea
\bea
|M(u_R g \to be^+v)|^2 &=& \frac{64}{\Lambda ^4}  g_2^2 v^2 f_R^2 p_g\cdot p_u \left(g_{W}^{ L~ 2}
   M_T^2 ~ k_e \cdot p_g k_{ve}\cdot
   k_b  + g_{W}^{ R~ 2}
   s ~ k_{ve} \cdot p_u k_{e}\cdot k_b   \right) \nonumber \\ && \cdot
 \frac{\pi   \delta ( k_W^2 -m_W^2 )}{\Gamma_W m_W}\frac{C_F \cdot N_c}{((s-M_T^2)^2 + \Gamma_T^2 M_T^2) }
\eea
where $C_F= 4/ 3 $ and $N_c = 3 $ are the color factors.  A symmetry exists  between the $f_L$ and $f_R$ scenarios that   $k_{ve}$ is exchanged with $k_{e} $ and  the (V-A) coupling $g_{W}^L $ is exchanged with the (V+A) coupling $g_{W}^R$.  Phrasing the final states phase space in the rest frame of the  heavy top quark,
the differential cross section is  expressed as:
\bea
d \hat \sigma = \frac{1}{2s} \frac{|M|^2}{4 \cdot 3 \cdot 8 }  \frac{4}{(4 \pi)^5}
d E_b d E_\ell d \cos \theta_\ell  d\phi_\ell d \phi_{b \ell}
\eea
with the following integration limits:
\bea
 \frac{ \sqrt s }{2} - E_b & <& E_\ell  <  \frac{\sqrt s}{2}   \nonumber \\
  0 &<& E_b < \frac{\sqrt s }{2}   ,  \nonumber \\
  -1< \cos \theta_\ell < 1,  &&  0  < \phi_\ell  , \phi_{b \ell} < 2 \pi
    \eea
We have used the narrow width approximation for the $W^+ $ gauge bosons when we are calculating the amplitudes squared  and the delta function
can be written as :
\bea \delta (k_W^2 - m_W^2) = \frac{1}{{2\sqrt s }}\delta \left( {{E_b} - \frac{{\sqrt s }}{2}\left( {1 - \frac{{m_W^2}}{s}} \right)} \right).
\eea
The $ \cos \theta_{b \ell} $ is not independent which can be written in terms of  $E_b $ and $E_\ell$ by the following expression:
\bea
\cos \theta_{b \ell} =  \frac {s  - 2 \sqrt s  ( E_b + E_\ell ) }{2 E_b E_\ell}  + 1
\eea
Now we put all the elements into the amplitudes squared and after integrating out  $E_b $, $E_\ell$  and two azimuthal angles of $ \phi_\ell $ and $\phi_{b \ell } $, we get
the following differential cross sections with respect to $\cos \theta_\ell$ which depict the angular distribution of the charged lepton in the center of mass rest  frame:
\bea
&&\frac{ d \hat \sigma(u_L g \to b\ell^+ \nu) }{d \cos \theta_\ell }  = \frac{   g_2^2 v^2 f_L^2  }{3^2 2^{11} \pi^2  \Lambda^4 }\frac{1}{\Gamma_W m_W} \frac{1}{s} \left(  g_{W}^{L ~2} \cdot s
   \left(m_W^2-s\right)^2 (2 m_W^2+s ) \cdot  (1+ \cos\theta_\ell)  \right.  \nonumber \\ && \qquad  -   \left.   g_{W}^{R ~2}  M_T^2  \left( \left(m_W^2-s\right)^2 (2 m_W^2+s ) +12 m_W^4 s -12 m_W^2  s^2 + 12 m_W^4 s \log \left[\frac{s}{m_W^2}\right] \right) \cos \theta_\ell   \right. \nonumber \\  &&  \qquad  +   \left.  g_{W}^{R ~2}  M_T^2  \left(m_W^2-s\right)^2 (2 m_W^2+s )  \right)  \cdot  \frac{1}{((s-M_T^2)^2 + \Gamma_T^2 ) }
 \eea
\bea
&& \frac{ d \hat \sigma(u_R g \to b\ell^+ \nu) }{d \cos \theta_\ell } = \frac{   g_2^2 v^2 f_R^2  }{3^2 2^{11} \pi^2  \Lambda^4 }\frac{1}{\Gamma_W m_W} \frac{1}{s} \left(  g_{W}^{L ~2} \cdot M_T^2
\left(m_W^2-s\right)^2 (2 m_W^2+s ) \cdot  (1- \cos\theta_\ell)  \right.  \nonumber \\  && \qquad +  \left.   g_{W}^{R ~2}  s  \left(\left(m_W^2-s\right)^2 (2 m_W^2+s ) +12 m_W^4 s -12 m_W^2  s^2 + 12 m_W^4 s \log \left[\frac{s}{m_W^2}\right] \right) \cos \theta_\ell   \right. \nonumber \\  && \qquad +   \left.  g_{W}^{R ~2}  s   \left(m_W^2-s\right)^2 (2 m_W^2+s ) \right)  \cdot  \frac{1}{((s-M_T^2)^2 + \Gamma_T^2 ) }
 \eea
 As we can see, only the term proportional to the (V-A) coupling has exactly  the $1\pm \cos \theta_\ell $ distribution and the term proportional to the (V+A) coupling is slightly corrected to be $ (1\mp \cos \theta_\ell ) \pm \mathcal{O} (m_W^2 / s ) \cos \theta_\ell  $ , the deviation can be ignored in the limit of $m_W \ll  \sqrt s $ . 

\newpage
\appendix

\end{document}